\documentclass{emulateapj}
\usepackage{natbib}
\usepackage{rotating}
\usepackage{mathtools}
\usepackage{tablefootnote}

\usepackage{graphicx}
\usepackage{amssymb}
\usepackage[usenames]{color}

\begin{document}

\title[Nucleosynthesis of Mo and Ru isotopes] 
{Production of Mo and Ru isotopes in neutrino-driven winds:
implications for solar abundances and presolar grains}

\author{J. Bliss}
\affil{Institut f\"ur Kernphysik, Technische Universit\"at Darmstadt, Schlossgartenstr. 2,
Darmstadt 64289, Germany}
\email{julia.bliss@physik.tu-darmstadt.de}

\author{A. Arcones}
\affil{Institut f\"ur Kernphysik, Technische Universit\"at Darmstadt, Schlossgartenstr. 2,
Darmstadt 64289, Germany\\
\and GSI Helmholtzzentrum f\"ur Schwerionenforschung GmbH, Planckstr. 1, Darmstadt 64291, Germany}
\email{almudena.arcones@physik.tu-darmstadt.de}

\author{Y.-Z. Qian}
\affil{School of Physics and Astronomy, University of Minnesota, Minneapolis, MN 55455, USA\\
\and Tsung-Dao Lee Institute, Shanghai 200240, China}
\email{qian@physics.umn.edu} 

\begin{abstract}
The origin of the so-called $p$-isotopes $^{92,94}\mathrm{Mo}$ and 
$^{96,98}\mathrm{Ru}$ in the solar system remains a mystery as
several astrophysical scenarios fail to account for them. In addition,
data on presolar silicon carbide grains of type X (SiC X) exhibit peculiar Mo
patterns, especially for $^{95,97}\mathrm{Mo}$.  
We examine production of Mo and Ru isotopes in neutrino-driven 
winds associated with core-collapse supernovae (CCSNe) over a wide
range of conditions. We find that proton-rich winds can make dominant
contributions to the solar abundance of $^{98}\mathrm{Ru}$, significant
contributions to those of $^{96}$Ru ($\lesssim 40\%$) and $^{92}$Mo 
($\lesssim 27\%$), and relatively minor contributions to that of $^{94}$Mo 
($\lesssim 14\%$). In contrast, neutron-rich winds make negligible contributions 
 to the solar abundances of $^{92,94}$Mo and cannot produce $^{96,98}$Ru. 
However, we show that some neutron-rich winds can account for 
the peculiar Mo patterns in SiC X grains.
Our results can be generalized if conditions similar to those studied here 
are also obtained for other types of ejecta in either CCSNe or neutron star 
mergers.
\end{abstract}
\maketitle


\section{Introduction}
\label{sec:introduction}

The isotopic abundances of the solar system obtained from meteoritic data 
(see e.g., \citealt{Lodders:2003}) played a crucial 
role in establishing the framework of the basic processes of nucleosynthesis that gave rise
to these abundances in particular and were responsible for the chemical evolution of the
universe in general. For elements heavier than Fe, it is well known that the major sources
for their solar abundances are the slow ($s$) and rapid ($r$) neutron-capture processes 
\citep{Burbidge.etal:1957, Cameron:1957}.
In a number of cases, the most proton-rich isotopes of an element cannot be made by
either of these processes and must be attributed to the so-called $p$-process (see e.g., 
\citealt{Meyer:1994,Arnould.Goriely:2003}). There have
been both observational and theoretical studies that strongly support low-mass
($\sim 1.5$--$3\,M_\odot$) stars during the asymptotic giant branch (AGB) stage of their
evolution as the site for the main $s$-process producing Sr and heavier elements (see e.g., 
\citealt{Kaeppeler.etal:2011}). It is
also well known that massive ($>10\,M_\odot$) stars during their pre-supernova evolution
can produce nuclei up to $^{88}$Sr through the weak $s$-process starting with the Fe in
their birth material (see e.g., \citealt{Raiteri.etal:1993, Pignatari.etal:2010}). 
The $p$-process is usually associated with $(\gamma,n)$ reactions
on the preexisting nuclei as the shock propagates through the outer shells of a massive 
star during its supernova explosion (see \citealt{Pignatari.etal:2016} for a review).
A kilonova powered by the decay of newly-synthesized $r$-process nuclei
in a binary neutron star merger was observed recently (e.g., \citealt{Smartt:2017, Kasen:2017}).
This observation demonstrates that mergers of two neutron stars or a neutron star and 
a black hole are important sites for the $r$-process (see e.g.,
\citealt{Freiburghaus.etal:1999,Goriely.etal:2011,Korobkin.etal:2012}).
Other sites (see e.g., \citealt{Woosley.Hoffman:1992,Banerjee.etal:2011,Nishimura.etal:2015})
associated with core-collapse supernovae (CCSNe) from massive stars have also been proposed 
and may play an important role in $r$-process 
enrichment at the earliest epochs (see e.g., \citealt{Qian.Wasserburg:2007,Qian:2014,Hansen.etal:2014}).

In connection with modeling CCSNe (see e.g., \citealt{Janka:2012} for a review), 
new mechanisms for producing the elements from
Zn to Ag with mass numbers $A\sim 64$--110 have been discovered 
\citep{Woosley.Hoffman:1992,Hoffman.etal:1996,Pruet.etal:2006,Frohlich.etal:2006}. 
These are associated with neutrino-driven winds from the proto-neutron star created 
in a CCSN. Depending on the electron fraction, entropy, and expansion time scale 
(see e.g., \citealt{Qian.Woosley:1996}), major production of some of the
above elements occurs in the wind (see e.g., 
\citealt{Witti.etal:1994,Hoffman.etal:1997,Arcones.Montes:2011, Arcones.Bliss:2014}). 
In these processes, $(n,\gamma)$, $(n,p)$, $(p,\gamma)$,
$(\alpha,\gamma)$, $(\alpha,n)$, $(\alpha,p)$ and their inverse reactions are all important 
\citep{Woosley.Hoffman:1992,Bliss.etal:2017},
in contrast to the dominance of neutron capture in both the $s$- and $r$-processes.
For proton-rich winds, $\bar\nu_e+p\to n+e^+$ can provide neutrons to break through 
the bottleneck nuclei with slow $\beta$-decay by efficient $(n,p)$ reactions. This
$\nu p$-process \citep{Pruet.etal:2006, Frohlich.etal:2006, Wanajo:2006} can produce 
nuclei up to $A\sim 110$ and perhaps even further for the most favorable conditions. 

As described above, a wide range of nuclei with $A\sim 64$--110 conventionally 
assigned to the $s$-, $r$-, and $p$-processes can be produced in neutrino-driven winds 
through very different mechanisms. The corresponding yield patterns are sensitive to
the conditions in the wind (see e.g., \citealt{Arcones.Bliss:2014,Bliss.etal:2018}) and usually distinct 
from the solar abundance pattern in this region. The isotopes $^{92,94}$Mo and 
$^{96,98}$Ru are commonly taken to be produced by the $p$-process only, but
$p$-process models have difficulty accounting for their solar abundances
(see e.g., \citealt{Meyer:1994,Arnould.Goriely:2003}).
In this paper, we explore a wide range of wind conditions to study the production of 
Mo and Ru isotopes and the implications for the solar abundances of $^{92,94}$Mo 
and $^{96,98}$Ru. Further,
in light of the peculiar Mo patterns, especially for $^{95,97}$Mo, and the associated
anomalies in Zr found in presolar silicon carbide grains of type X (SiC X;
\citealt{Pellin.etal:1999,Pellin.etal:2006}), we also discuss possible nucleosynthetic
contributions to these grains from neutrino-driven winds. Our results can be
generalized if conditions similar to those explored here are also obtained for other
types of ejecta in either CCSNe or neutron star mergers. Consequently, our
study is complementary to post-processing studies based on specific simulations 
of these events.

\section{Parametric Models of Neutrino-Driven Winds}
\label{sec:method}

Nucleosynthesis in an expanding mass element starting
from high temperature and density depends on
the entropy $S$, expansion time scale $\tau$, and electron fraction
$Y_{e}$ \citep{Qian.Woosley:1996,Hoffman.etal:1997,Meyer:1997,
Freiburghaus:1999,Otsuki.etal:2000,Thompson.etal:2001}. So the main features of
such nucleosynthesis can be captured by parametric studies.
Typically the evolution of the temperature $T$ with time $t$ is taken 
as some function $T(t)$ characterized by the time scale $\tau$ [e.g., 
$T(t)\propto\exp(-t/\tau)$]. As the entropy $S$ is mainly a function of 
$T$ and the density $\rho$ (e.g., $S\propto T^3/\rho$ for 
radiation-dominated conditions), the time evolution of $\rho$ can be 
obtained from $T(t)$ by assuming conservation of $S$. For an initial
temperature $T(0)\sim 10$~GK, the nuclei present (predominantly
free neutrons and protons) are in nuclear statistical equilibrium (NSE), 
so the initial composition can be determined by the NSE equations 
along with mass and charge conservation. Because the matter of 
concern is neutral, the sum of all nuclear charge can be specified by 
the electron fraction $Y_e$. Thus, for a set of $S$, $\tau$, and $Y_e$, the 
corresponding nucleosynthesis can be followed with a reaction network. 

Compared to the generic parametric studies described above,
the studies presented here include additional ingredients of
realistic astrophysical environments with intense neutrino fluxes.
Specifically, we use as the baseline models the trajectories of three 
mass elements in the neutrino-driven wind from the proto-neutron
star in the CCSN model M15l1rl of
\cite{Arcones.etal:2007}. This explosion model has an efficient
neutrino transport scheme that allows to study the evolution of the 
wind for various progenitors in one and two dimensions
\citep{Arcones.etal:2007, Arcones.Janka:2011}. However,
simplifications in the treatment of neutrinos and the proto-neutron 
star lead to uncertainties in the explosion and the
subsequent neutrino-driven wind. In view of these uncertainties,
while we adopt the time evolution of the temperature $T(t)$ and the
radius $r(t)$ for three trajectories of wind mass elements ejected 
at time post core bounce $t_{\rm pb}=2$, 5, and 8~s, respectively,
we vary the entropy and the electron fraction.
For a trajectory with an original entropy $S_0$ and the corresponding
density evolution $\rho_0(t)$, we change the entropy to 
$S=(0.5$--$1.5)S_0$ and obtain the new density 
evolution $\rho(t)=\rho_0(t)S_0/S$ assuming $S\propto T^3/\rho$.
In addition, we vary the initial $Y_e$ over the range $Y_e(0)=0.45$--0.62 and
follow the subsequent evolution of $Y_e$ by including $\nu_e$
and $\bar\nu_e$ absorption on free neutrons and protons, respectively.
We assume that the $\nu_e$ ($\bar\nu_e$) spectrum is
Fermi-Dirac with temperature  $T_{\nu_e}$ ($T_{\bar\nu_e}$) and
zero chemical potential. Further, we fix $L_{\nu_e}=2\times 10^{51}$~ergs/s
and calculate $L_{\bar\nu_e}=L_{\nu_e}T_{\bar\nu_e}/T_{\nu_e}$ assuming
equal $\nu_e$ and $\bar\nu_e$ number fluxes.
For neutron-rich (proton-rich) winds, we fix $T_{\nu_e}=4$~MeV
($T_{\bar\nu_e}=4$~MeV) and choose $T_{\bar\nu_e}$ ($T_{\nu_e}$) 
to match the $Y_e(0)$. The $\nu_e$ and $\bar\nu_e$ fluxes experienced by a mass element
decrease with time as $r(t)^{-2}$. The parametric models for the neutrino-driven
wind described above should represent rather well both the variation of 
conditions with the time of ejection from the proto-neutron star in an individual 
CCSN and the variation among CCSNe from different progenitors.

\begin{figure}[h!]
\centering
\includegraphics[width=0.85\linewidth]{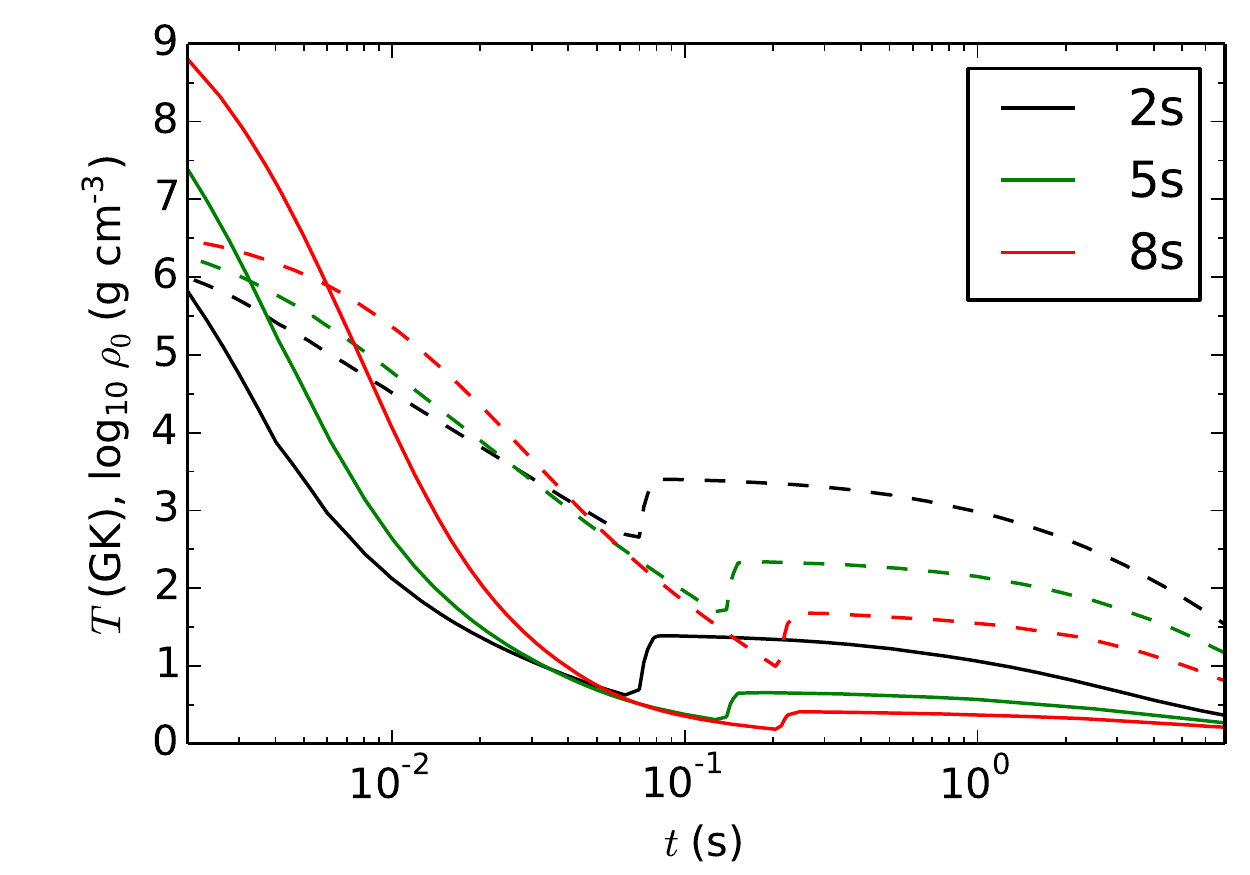}\\
\includegraphics[width=0.85\linewidth]{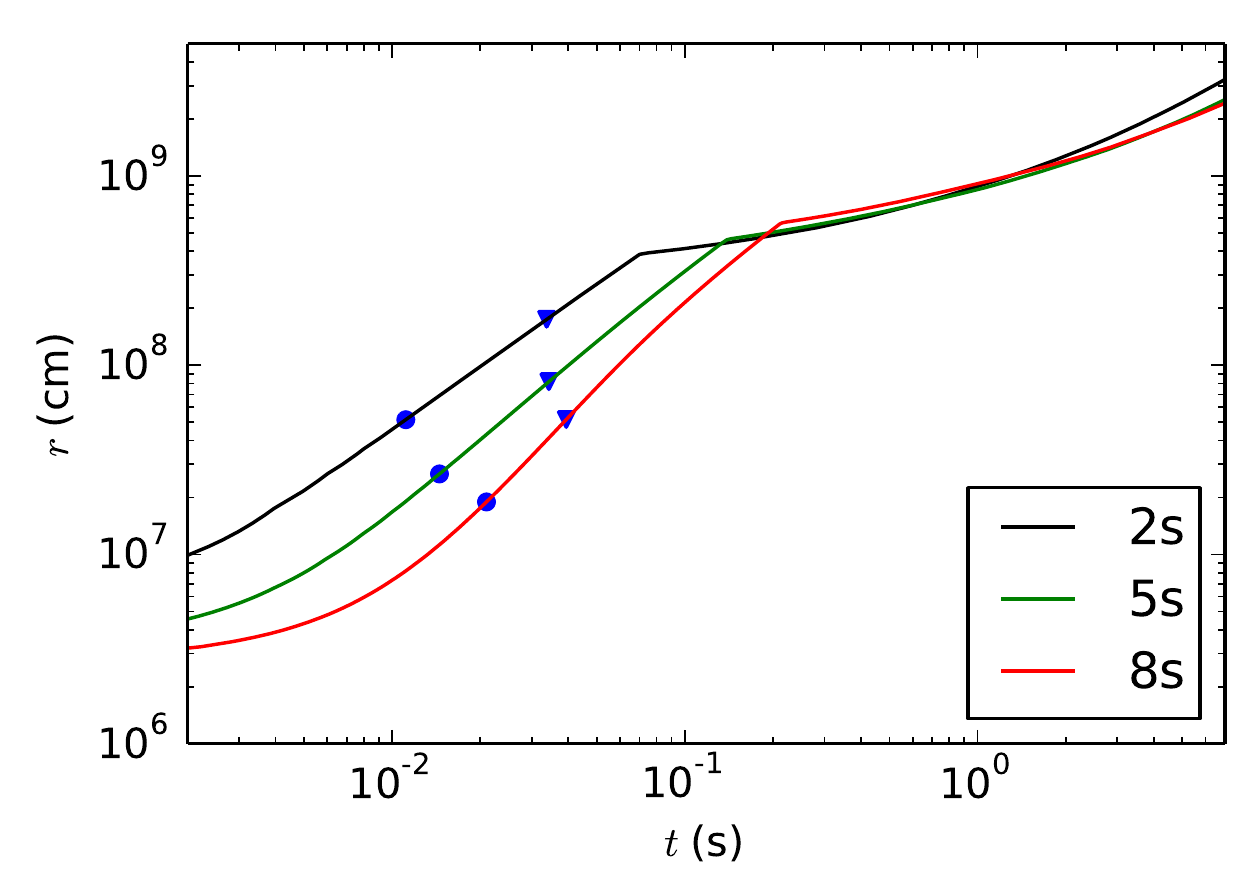}\\
\includegraphics[width=0.85\linewidth]{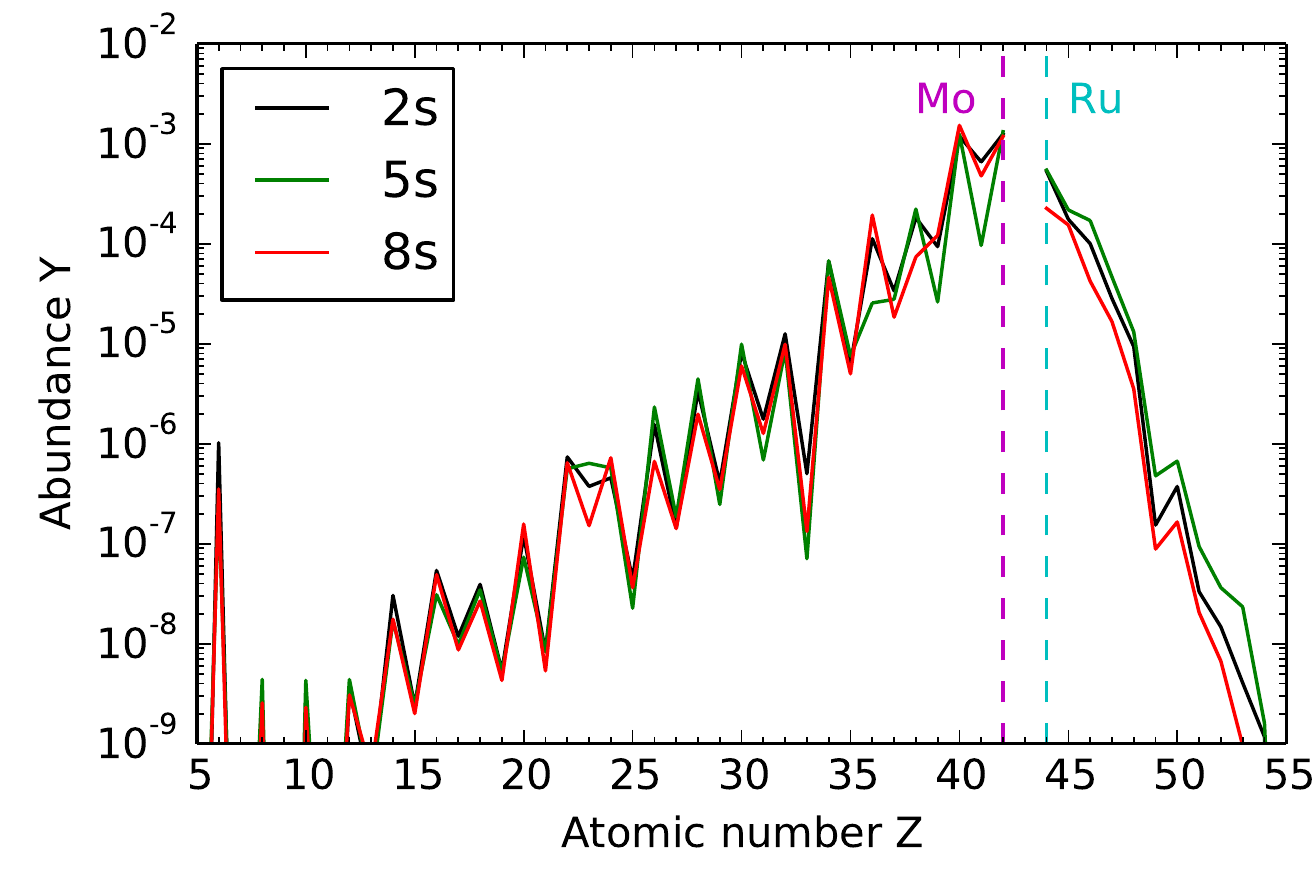}\\
\includegraphics[width=0.85\linewidth]{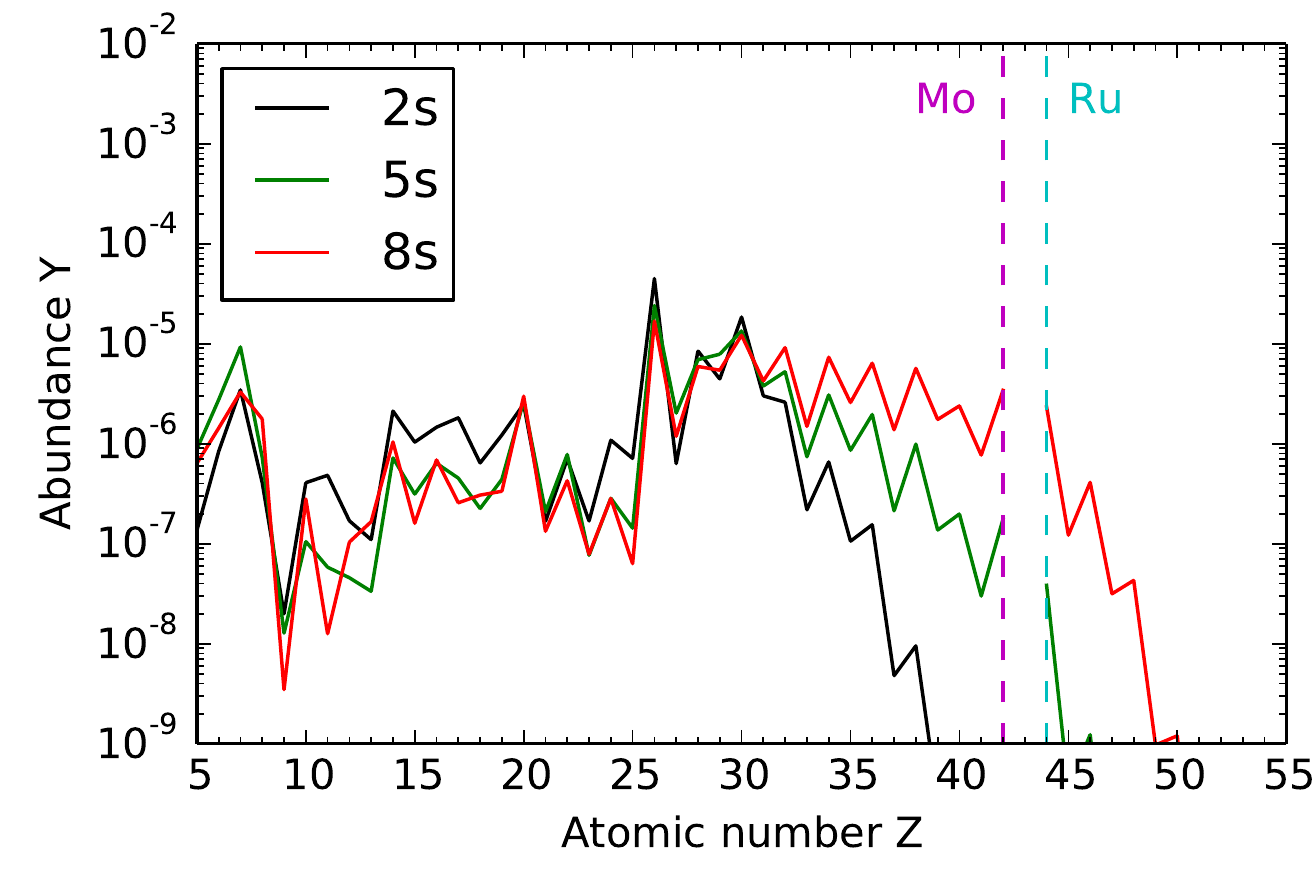}
\caption{Baseline models for nucleosynthesis. (a) Time evolution of 
$T/{\rm GK}$ (solid curves) and $\log(\rho/{\rm g\ cm}^{-3})$ 
(dashed curves) for trajectories of wind mass elements ejected 
at $t_{\rm pb}=2$ (black curves), 5 (green curves), and 8~s (red curves), 
respectively. (b) Time evolution of the radius $r$ for each trajectory.
The region of $T=2$ to 1~GK important to the $\nu p$-process is
indicated as that between the solid circle and triangle. 
(c) The final elemental (number) abundances for
each trajectory assuming $Y_e(0)=0.45$ (neutron-rich).
(d) Same as (c), but for $Y_e(0)=0.60$ (proton-rich).}
\label{fig:trajectories}
\end{figure}

  In Fig.~\ref{fig:trajectories}a, we present the baseline temperature 
  and density evolution for the three chosen trajectories. The baseline
  entropy is $S_0=67$, 78, and 85 in units of Boltzmann constant per nucleon
  ($k_\mathrm{B}/$nuc)
  for the wind mass element ejected at $t_{\rm pb}=2$, 5, and 8~s,
  respectively. The kinks in $T(t)$ and $\rho_0(t)$ are caused by the
  reverse shock when the wind runs into the material ejected at
  earlier times. Clearly, this type of time evolution cannot be simply
  parametrized by a single time scale $\tau$. As discussed below,
  the reverse shock at wind termination plays a crucial role in the
  $\nu p$-process (see e.g., \citealt{Wanajo.etal:2011,Arcones.etal:2012}).
  
  To calculate the nucleosynthesis, we use the same reaction network as
  in \cite{Frohlich.etal:2006}, which includes 4053 nuclei corresponding to
  the elements from H to Hf.  
  The reaction rates are taken from JINA ReaclibV1.0 \citep{Reaclib},
  which is a compilation of theoretical rates from 
  \cite{Rauscher.Thielemann:2000} and experimental rates from 
  \cite{Angulo.etal:1999}. The theoretical weak reaction rates in 
  \cite{Frohlich.etal:2006} are supplemented with experimental
  $\beta$-decay rates \citep{NuDat2} when available. The
  calculations start at $T(0) \sim 10$~GK, for which the composition
  is calculated from NSE for the specific $Y_{e}(0)$. The subsequent 
  evolution of the nuclear composition and $Y_e$ is calculated using the 
  full network that includes $\nu_e$ and $\bar\nu_e$ absorption on
  free neutrons and protons, respectively \citep{Frohlich.etal:2006}. 
  As described above, we always adjust the $\nu_e$ and $\bar\nu_e$ 
  luminosities and spectra so that they are consistent with the specified 
  $Y_e(0)$ \citep{Arcones.Bliss:2014}. The evolution of the radius
  $r(t)$ used to calculate that of the $\nu_e$ and $\bar\nu_e$ fluxes is
  shown in Fig.~\ref{fig:trajectories}b.

  In Figs.~\ref{fig:trajectories}c and \ref{fig:trajectories}d, 
  we illustrate the nucleosynthesis for our baseline models 
  assuming $Y_e(0)=0.45$ (neutron-rich)
  and 0.60 (proton-rich), respectively. It can be seen that
  the variations of $T(t)$ and $S_0$ among the three trajectories
  have a minor impact on the final abundances for neutron-rich 
  conditions when the same $Y_e(0)$ is used.
  However, the situation is markedly different for proton-rich
  conditions, where increasingly heavier nuclei are produced
  for the wind mass elements ejected at later times. This 
  is because the expansion affects the $\nu p$-process in such conditions 
  through the determination of both the number ratio of protons 
  to seed nuclei and the $\bar\nu_e$ flux at the onset of the process.
  Specifically, the expansion time scale for $T \sim 6$ to 3~GK sets the 
  number ratio of protons to seed nuclei at the onset of the $\nu p$-process. 
  In addition, this time scale also determines the radius of the mass element,
  and hence the $\bar\nu_e$ flux it receives, at this onset. Finally, the 
  combination of this $\bar\nu_e$ flux and the expansion time scale for 
  $T \sim 2$ to 1~GK determines the extent of neutron production by
  $\bar\nu_e$ absorption on protons during the $\nu p$-process. Clearly, 
  a sufficient neutron abundance is required to overcome the bottlenecks 
  on the path to heavier nuclei via $(n,p)$ reactions. The slowing down of
  the mass element near the wind termination helps to fulfill this requirement.
  The trajectory ejected at $t_{\rm pb}=2$~s has the fastest initial expansion
  and the mass element is already at a large radius at the onset of the
  $\nu p$-process (see Fig.~\ref{fig:trajectories}b). Therefore, for this
  trajectory $\bar\nu_e$ absorption on protons is strongly reduced during
  the $\nu p$-process even with the slowing down of the expansion
  near the wind termination. In contrast, the overall expansion for the trajectory 
  ejected at $t_{\rm pb}=8$~s is slower so that the $\nu p$-process becomes 
  much more efficient and reaches significantly heavier nuclei.
  
In the following discussion, we will focus on the
trajectory ejected at $t_{\rm pb}=8$~s. Pertinent results for
the other two trajectories are also presented. As described above,
our study is parametric in that we adopt fixed evolution of temperature
$T(t)$ and radius $r(t)$ for a wind trajectory while varying $S$ and 
$Y_e(0)$ to obtain a range of $\rho(t)$ and $Y_e(t)$, respectively. 
We note that only accurate modeling 
of a specific astrophysical environment can yield a self-consistent set
of $T(t)$, $\rho(t)$, and $Y_e(t)$. In the case of the neutrino-driven
wind, this would require accurate neutrino transport in the
proto-neutron star and accurate simulation of the CCSN
explosion \citep{Arcones.Thielemann:2012,Arcones.Bliss:2014}. 
While our parametric approach can only serve as approximation to 
the rigorous astrophysical models, it captures the salient features
of such models that are important to heavy-element nucleosynthesis.
Further, it is efficient for surveying a wide range of possibilities.
The results from our parametric models should provide good guidance 
in finding the conditions for producing Mo and Ru isotopes of interest.
 
\section{Results on Mo and Ru Isotopes}
\label{sec:results}
\begin{figure*}[!tb]
\centering
    \includegraphics[width=0.43\textwidth]{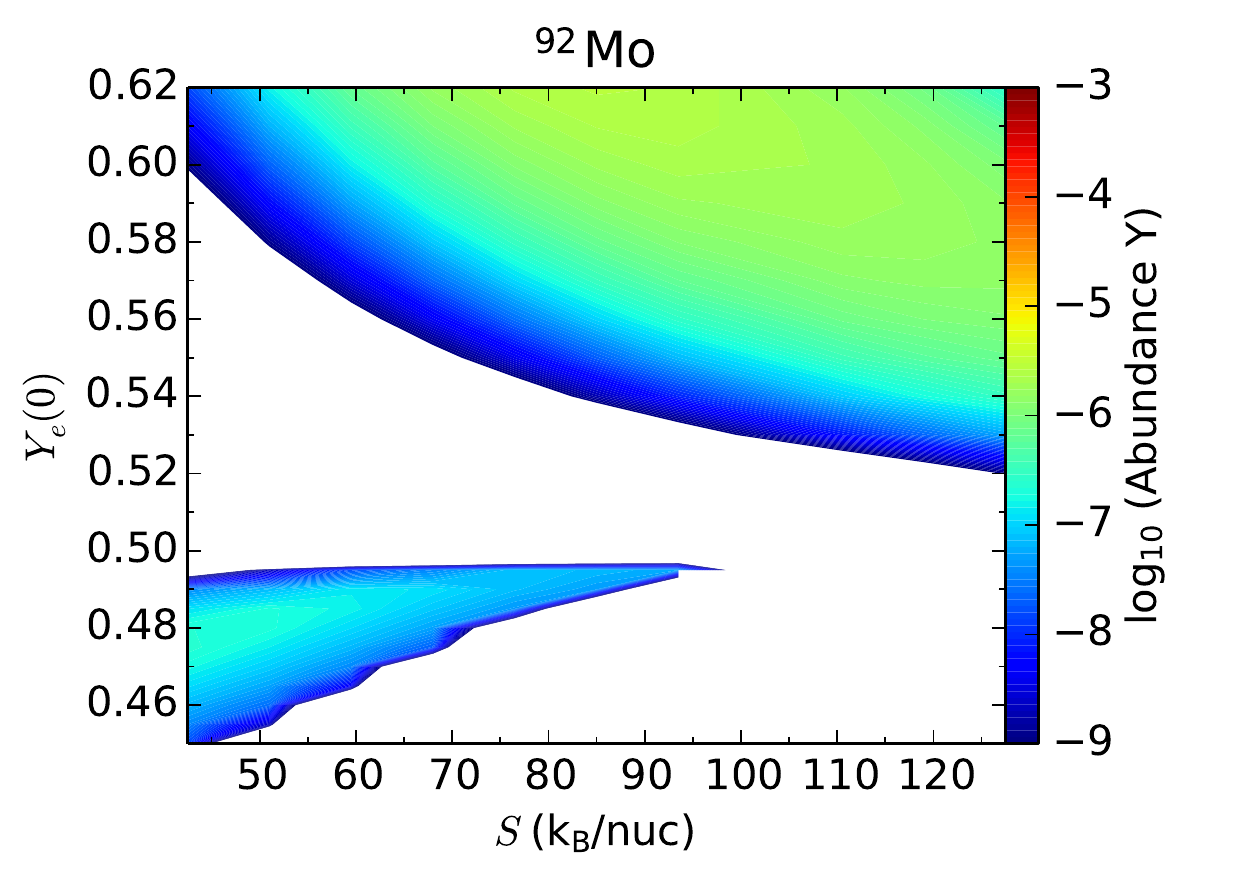}%
    \includegraphics[width=0.43\textwidth]{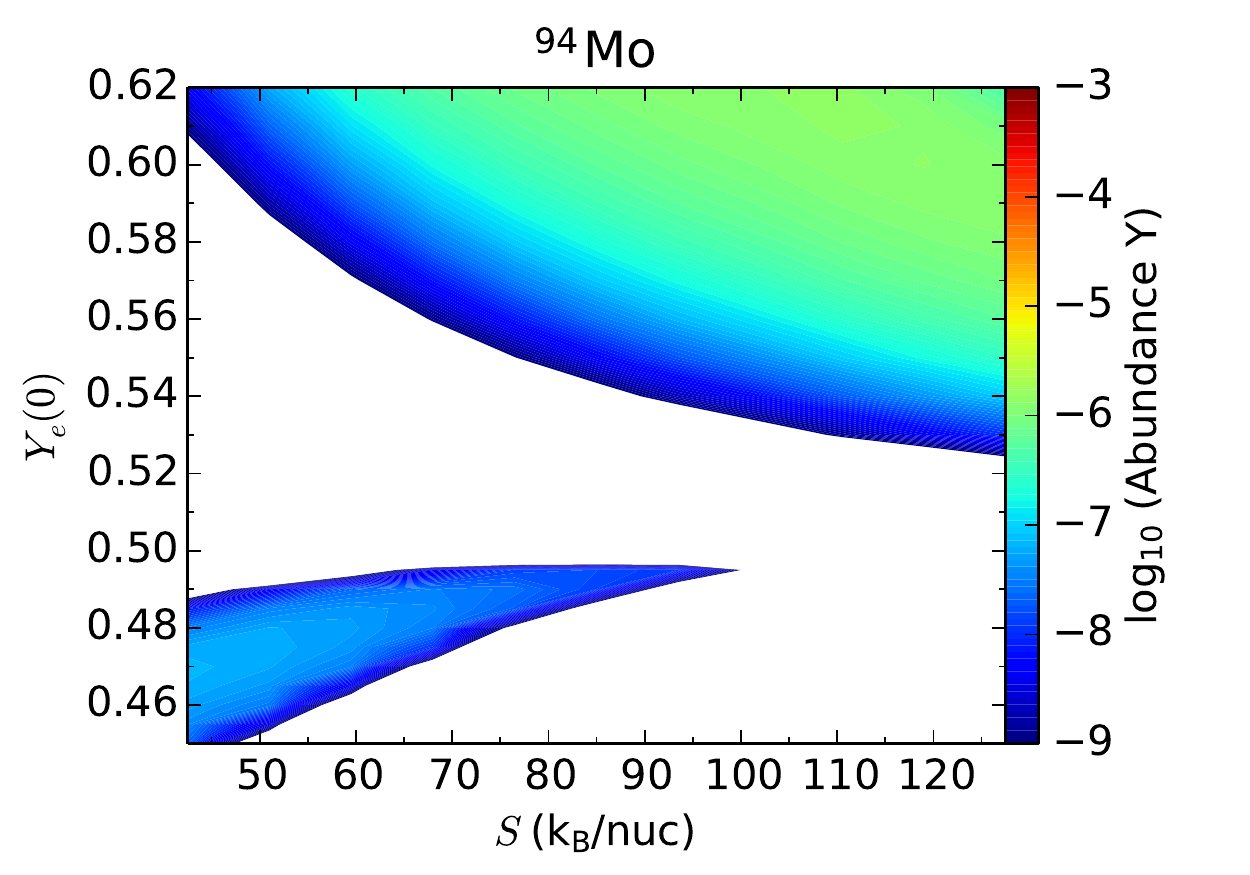}\\
    \includegraphics[width=0.43\textwidth]{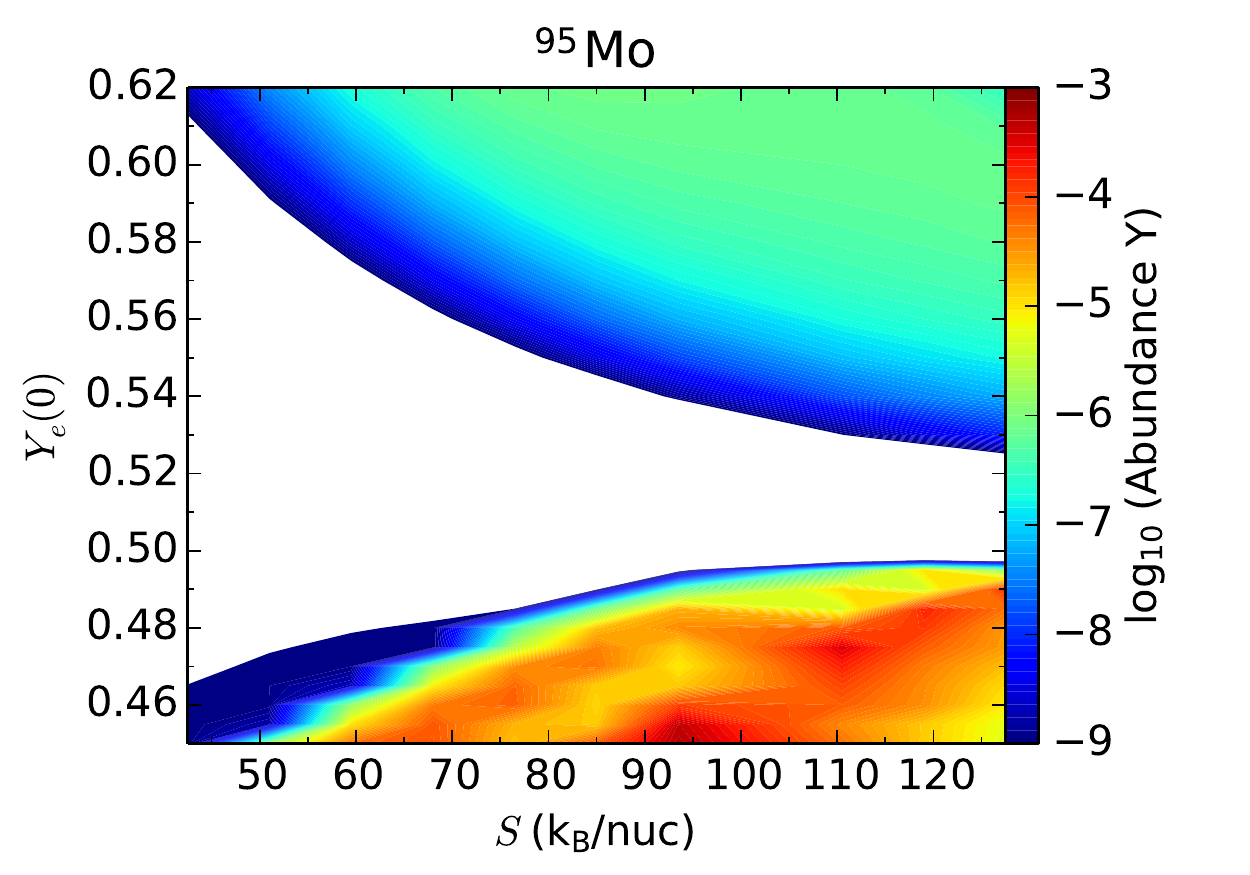}%
    \includegraphics[width=0.43\textwidth]{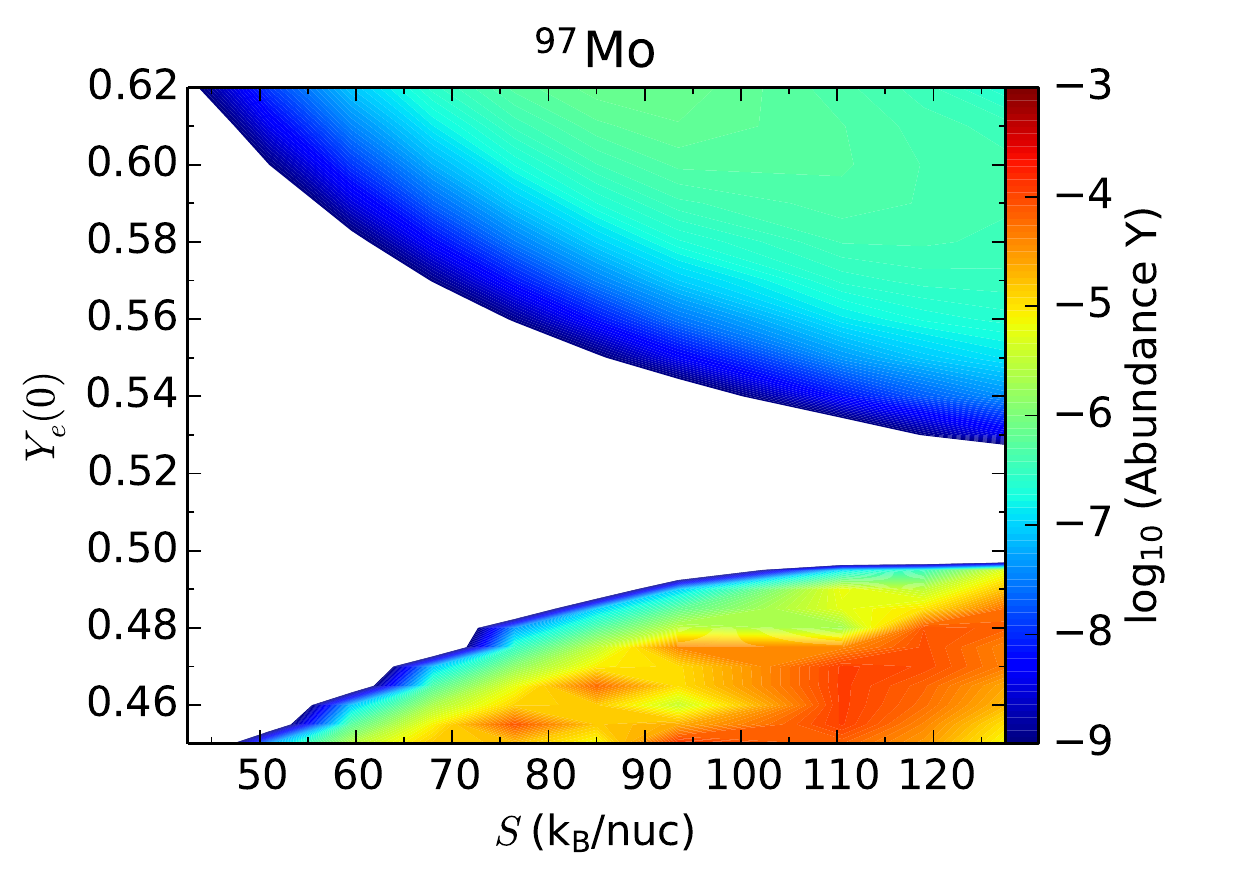}\\ 
    \includegraphics[width=0.43\textwidth]{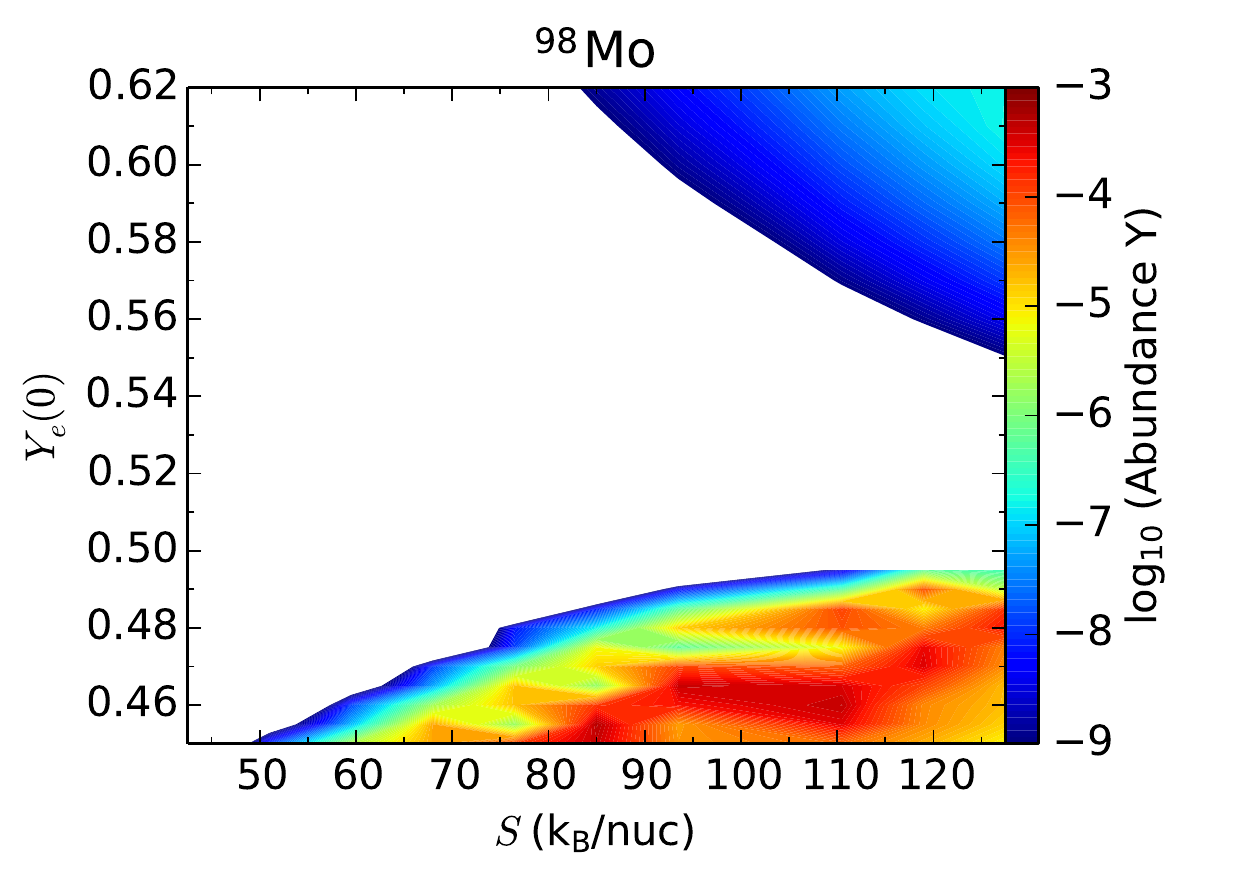}%
    \includegraphics[width=0.43\textwidth]{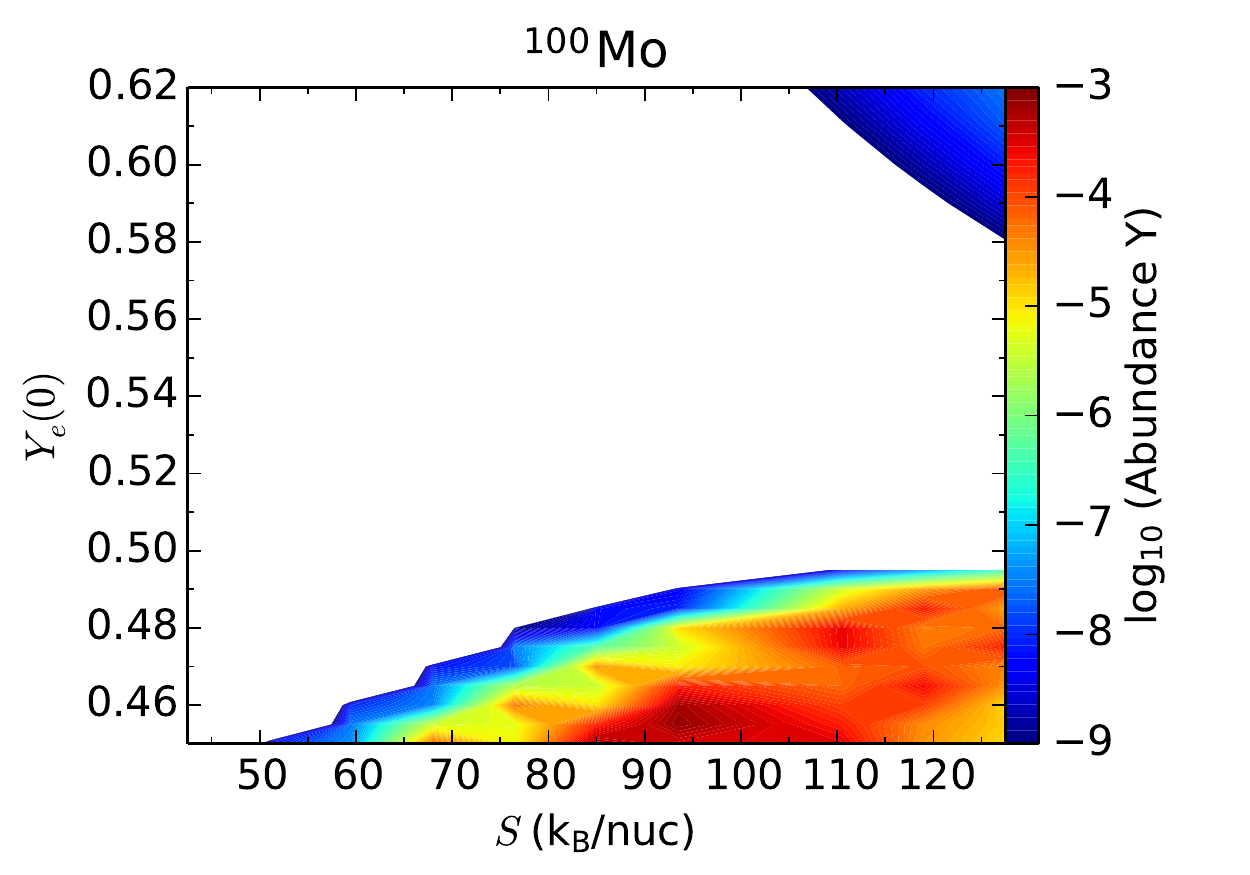}
    \caption{Color-coded contours illustrate the abundances (on a logarithmic scale)
    of the indicated isotope produced by the trajectory ejected at $t_{\rm pb}=8$~s
    for different $Y_{e}(0)$ and $S$. Abundances smaller than $10^{-9}$ are shown in
      white.}
\label{fig:ab_Mo}
\end{figure*}

\begin{figure*}[!tb]
\centering
\includegraphics[width=0.43\textwidth]{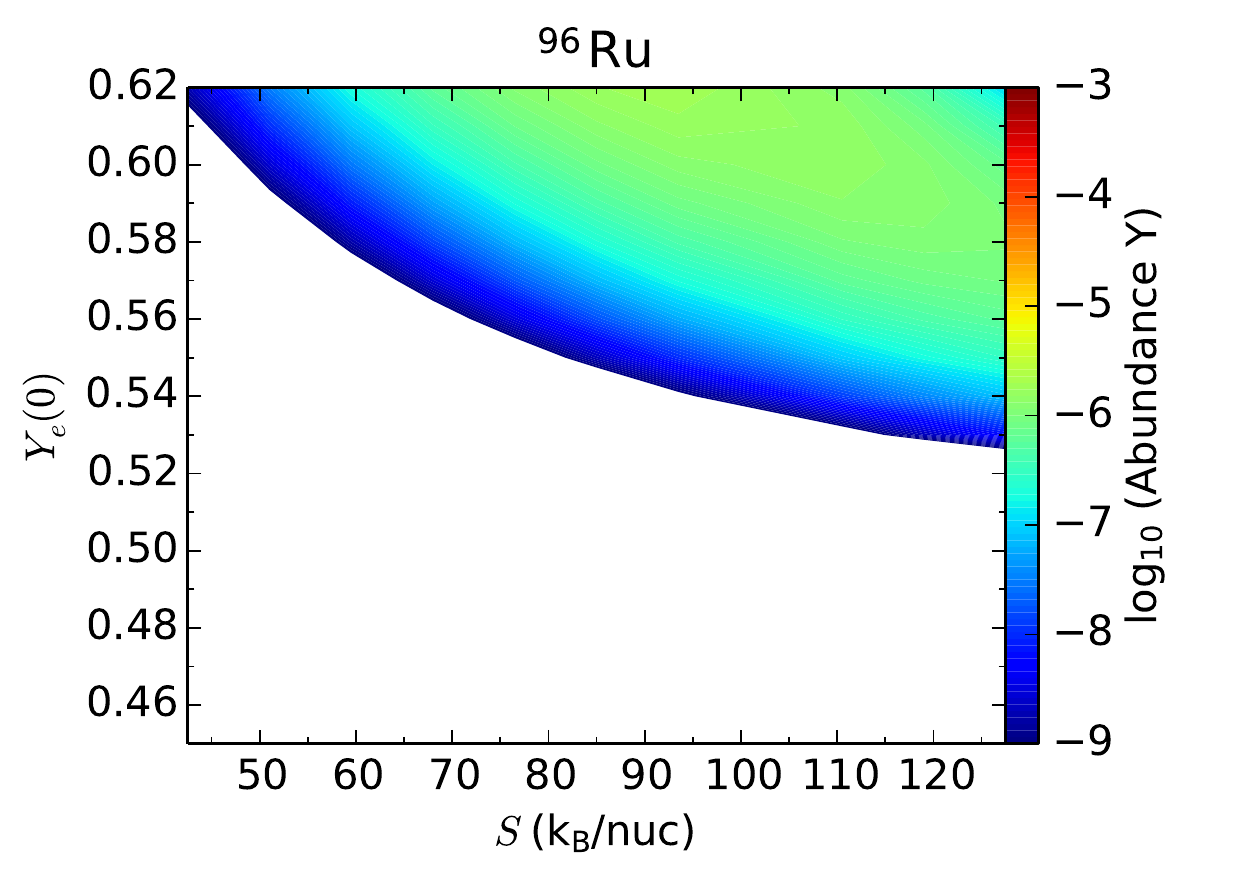}%
\includegraphics[width=0.43\textwidth]{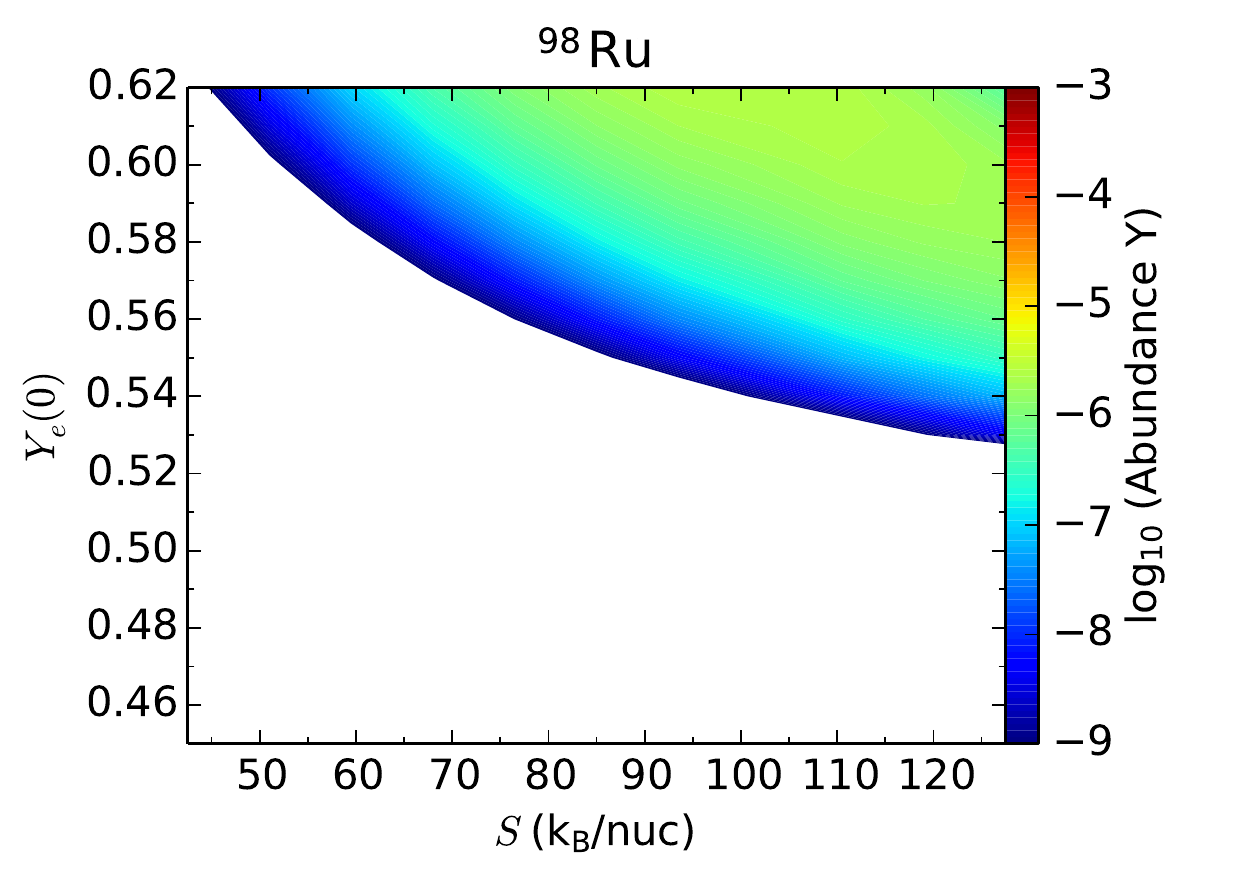}\\
\includegraphics[width=0.43\textwidth]{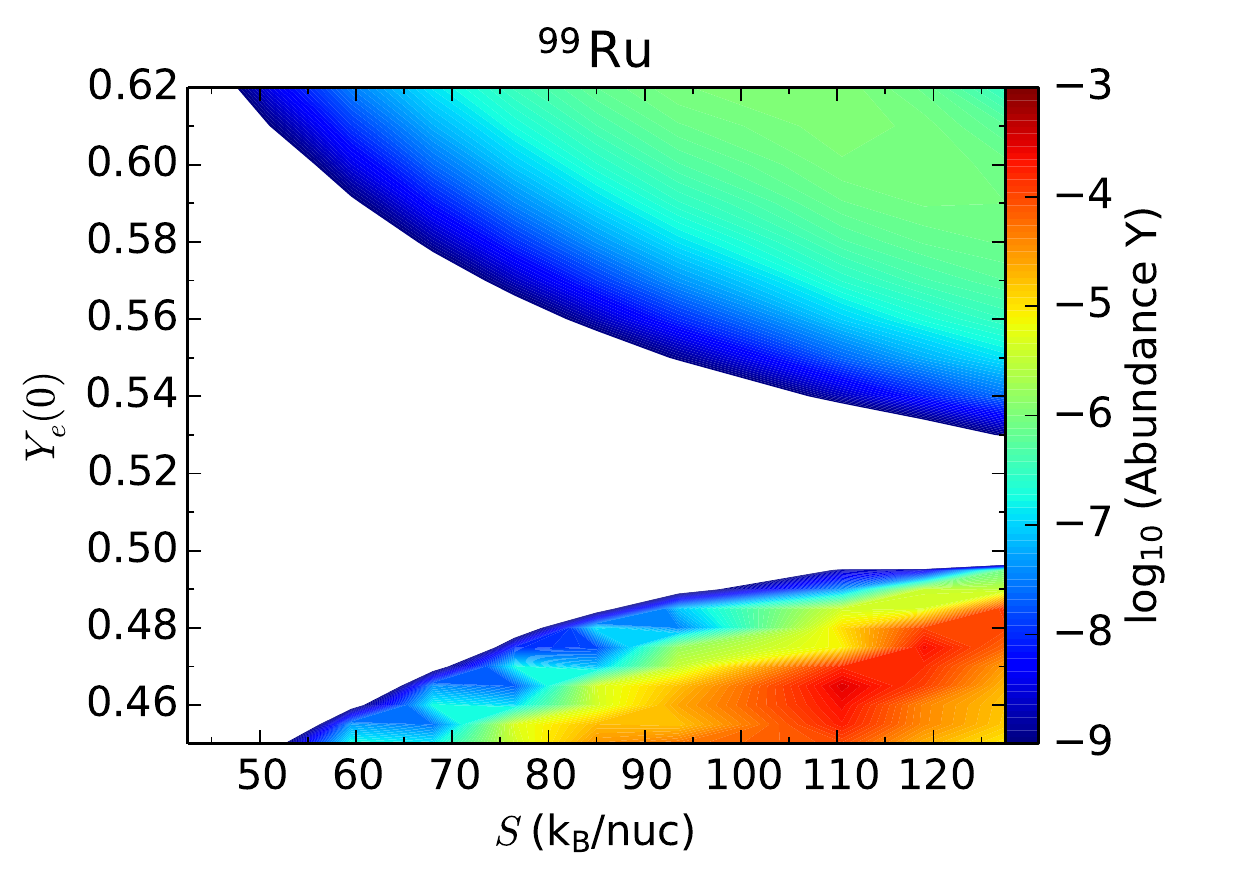}%
\includegraphics[width=0.43\textwidth]{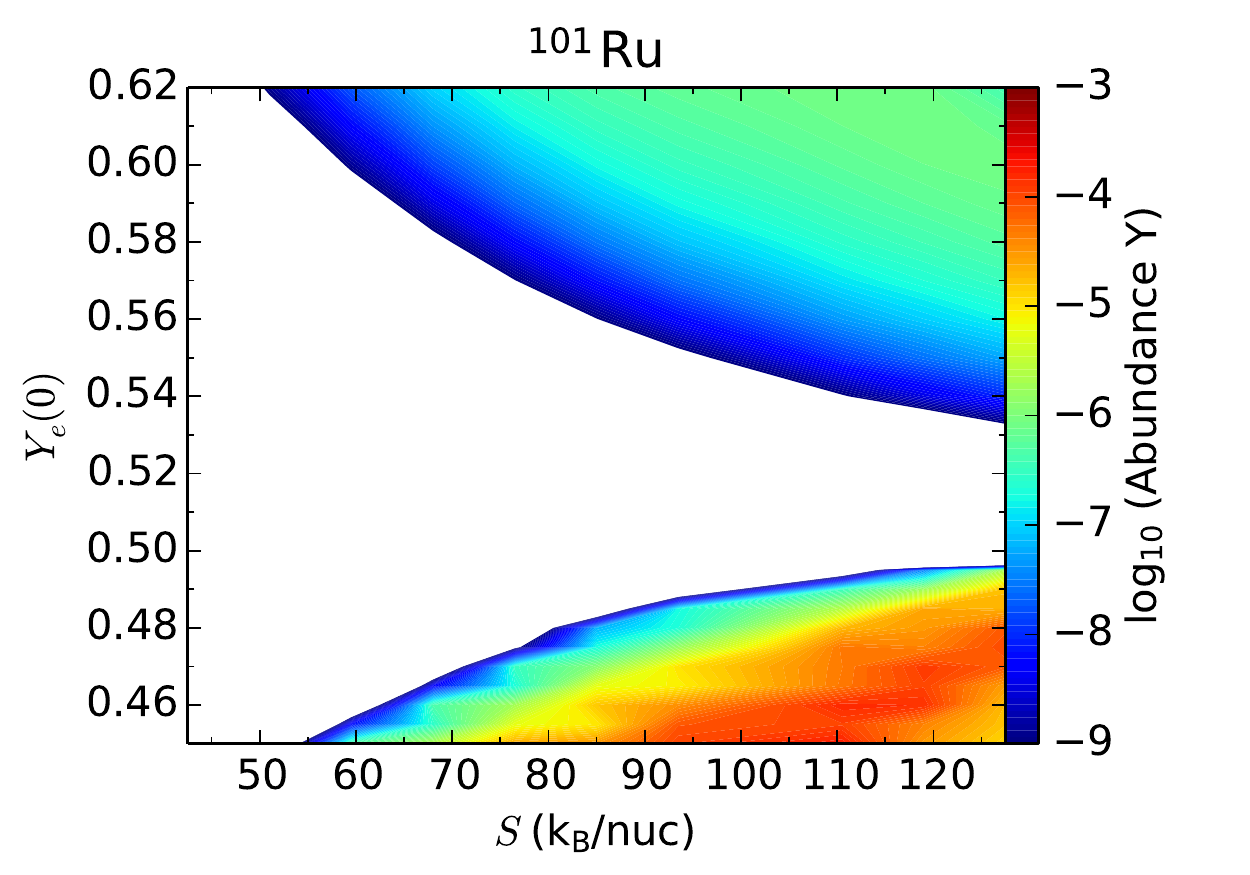}\\
\includegraphics[width=0.43\textwidth]{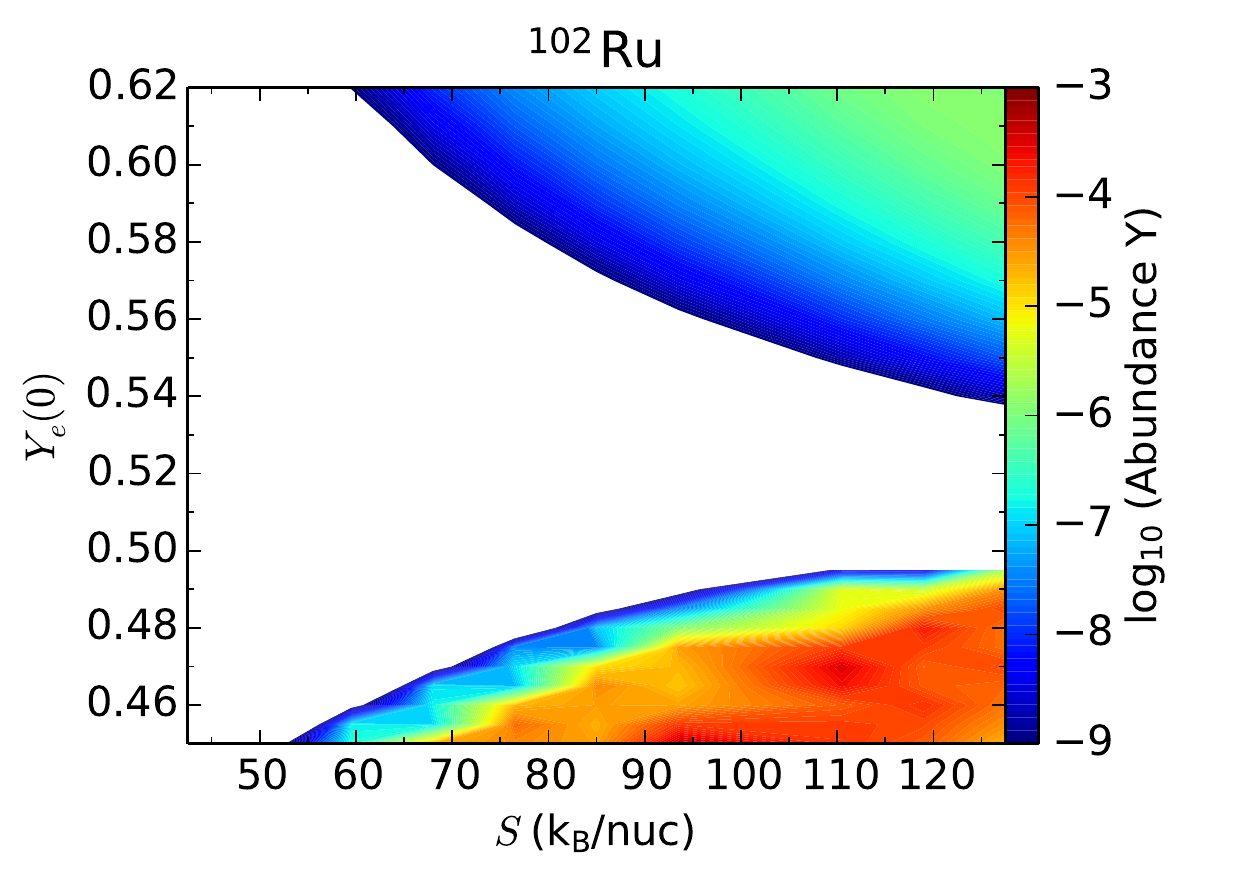}%
\includegraphics[width=0.43\textwidth]{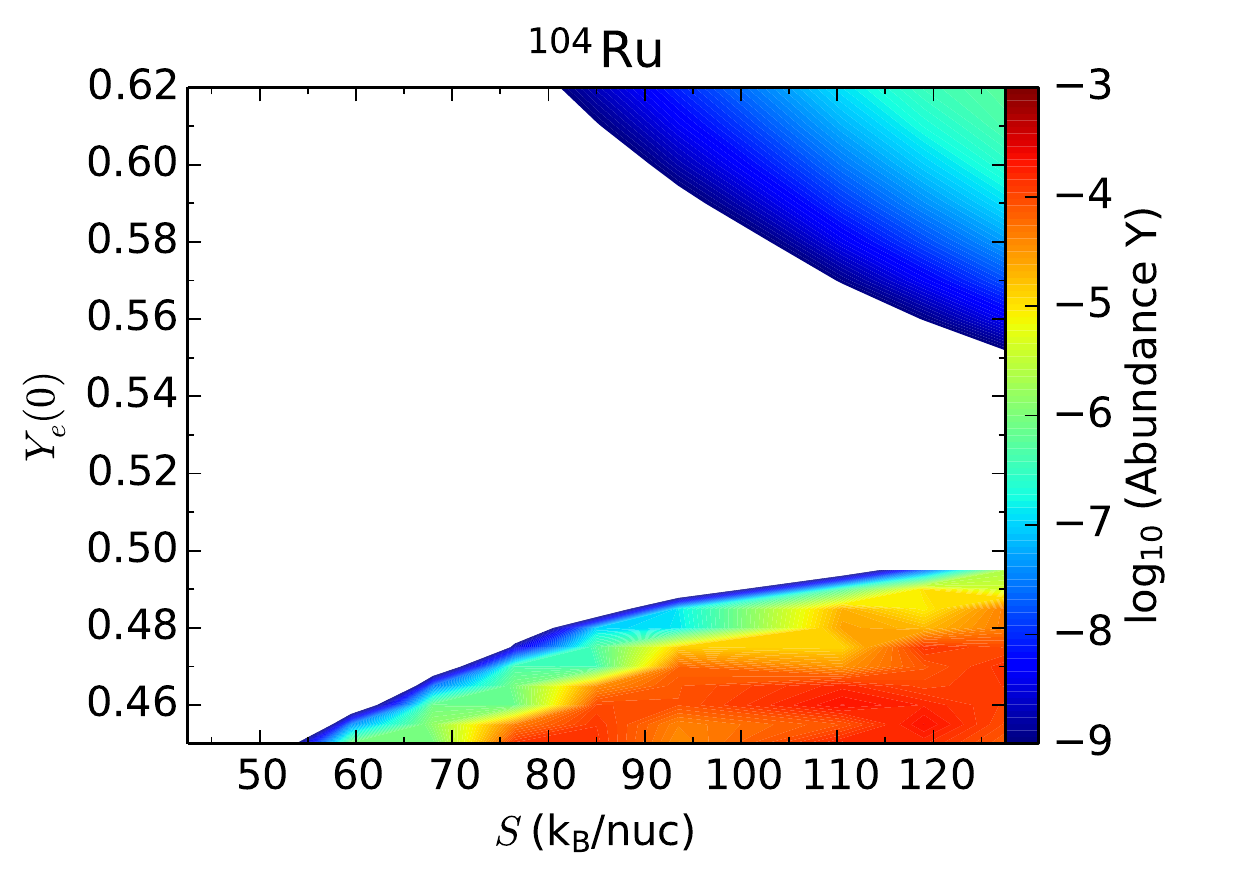}
    \caption{Same as Fig.~\ref{fig:ab_Mo}, but for Ru isotopes.}
\label{fig:ab_Ru}
\end{figure*}  
 
We are interested in the conditions under which various Mo and Ru 
isotopes can be made in environments similar to the neutrino-driven wind.  
In conventional terms, the seven stable isotopes of Mo (Ru) fall into
four categories: (1) $^{92,94}\mathrm{Mo}$ ($^{96,98}\mathrm{Ru}$)
as the $p$-only nuclei, (2) $^{96}\mathrm{Mo}$
($^{100}\mathrm{Ru}$) as the $s$-only nuclide,
(3) $^{95,97,98}\mathrm{Mo}$ ($^{99,101,102}\mathrm{Ru}$) as the
mixed nuclei with contributions from the $s$- and $r$-processes,
and (4) $^{100}\mathrm{Mo}$ ($^{104}\mathrm{Ru}$) as the $r$-only
nuclide. As mentioned in the introduction,
all these isotopes can be produced in the wind by processes that
differ from the conventional $p$-, $s$-, and $r$-processes. However,
for convenience, we will refer to these isotopes using the above
conventional category labels.

Based on the trajectory ejected at $t_{\rm pb}=8$~s,
we present in Fig.~\ref{fig:ab_Mo} (\ref{fig:ab_Ru}) the abundances
of various Mo (Ru) isotopes as functions of $S$ and
$Y_e(0)$. Figure~\ref{fig:ab_Mo} shows that both 
the $p$-only nuclei $^{92,94}\mathrm{Mo}$ can be produced 
in slightly neutron-rich [$Y_e(0)<0.5$] or proton-rich [$Y_e(0)>0.5$] winds 
for a significant range of $S$.
In contrast, the $p$-only nuclei $^{96,98}\mathrm{Ru}$ can be produced
only in proton-rich winds but not in neutron-rich winds 
(see Fig.~\ref{fig:ab_Ru}).

 Two distinct processes are responsible for producing 
 $^{92,94}\mathrm{Mo}$ in neutron- and proton-rich
 conditions, respectively. Neither of these processes 
 are dominated by neutron capture. To reach $^{92,94}\mathrm{Mo}$ 
 in neutron-rich conditions, the nucleosynthesis path has to move along 
 the valley of stability. If the neutron abundance is too large, the 
 path goes farther to the neutron-rich side and production of
 $^{92,94}\mathrm{Mo}$ is blocked by $^{92,94}\mathrm{Zr}$, 
 respectively. Therefore, production of $^{92,94}\mathrm{Mo}$ 
 requires that very few neutrons be present at freeze-out of
 charged-particle reactions with the composition dominated by 
 $\alpha$-particles and seed nuclei (see also 
 \citealt{Hoffman.etal:1996, Wanajo:2006,Farouqi.et.al.2009}).
 In addition to neutron capture, reactions involving light 
 charged-particles (e.g., protons and $\alpha$-particles) play 
 important roles in producing the seed nuclei including
 $^{92,94}\mathrm{Mo}$. Larger abundances of these two
 isotopes are obtained for smaller $S$, which favors a smaller
 number ratio of neutrons to the seed nuclei.
 
 In proton-rich conditions, $^{92,94}\mathrm{Mo}$ are produced
 by the $\nu p$-process, where $(p,\gamma)$ and $(n,p)$ reactions 
 play key roles in driving the flow towards heavy nuclei. The
 $(p,\gamma)$ reactions clearly depend on the proton abundance, 
 and so do the $(n,p)$ reactions with
 $\bar\nu_e+p\to n+e^+$ providing the neutrons. 
 Because a larger $Y_{e}(0)$ and a higher $S$
 favor a higher proton abundance, the $\nu p$-process produces
 more $^{92,94}\mathrm{Mo}$ with increasing $Y_{e}(0)$ and $S$ 
 initially \citep{Pruet.etal:2006}. However, when the number
 ratio of protons to the seed nuclei reaches a threshold, the 
 abundances of $^{92,94}\mathrm{Mo}$ start to decrease
 as the flow moves towards heavier nuclei.
 
The abundances of $^{92,94}$Mo are rather low in neutron-rich
 conditions (see Fig.~\ref{fig:ab_Mo}).
 Because the nucleosynthesis path towards the $p$-only nuclei 
 $^{96,98}\mathrm{Ru}$ passes through $^{92,94}$Mo, no
 significant amounts of $^{96,98}\mathrm{Ru}$ can be synthesized
 in neutron-rich conditions. In contrast,
 $^{96,98}\mathrm{Ru}$ can be produced in proton-rich conditions
 similarly to $^{92,94}\mathrm{Mo}$ (see Fig.~\ref{fig:ab_Ru}).
 


The mixed nuclei $^{95,97,98}\mathrm{Mo}$ and
$^{99,101,102}\mathrm{Ru}$, as well as the $r$-only nuclei 
$^{100}\mathrm{Mo}$ and $^{104}\mathrm{Ru}$, 
are much more abundant in neutron-rich conditions
(see Figs.~\ref{fig:ab_Mo} and \ref{fig:ab_Ru}). 
Their production is due to neutron capture and generally 
increases for lower $Y_e(0)$ and higher $S$, which favor a higher 
ratio of neutrons to the seed nuclei. For typical neutron-rich 
conditions investigated here, the neutron-capture process stays 
close to the valley of stability on the neutron-rich side. Once the 
neutrons are consumed, $\beta$-decays populate 
$^{95,97,98,100}\mathrm{Mo}$ and $^{99,101,102,104}\mathrm{Ru}$.
For conditions giving rise to a high ratio of
neutrons to the seed nuclei, the nucleosynthesis path moves
further away from stability and towards heavier nuclei, which
leads to a decrease in production of Mo and Ru isotopes.

The $s$-only nuclei $^{96}\mathrm{Mo}$ and $^{100}\mathrm{Ru}$ 
can be synthesized only in proton-rich winds by late $(n, \gamma)$
reactions (see~\citealt{Arcones.etal:2012,Frohlich.etal:2006,Wanajo:2006}). 
For this production channel to occur, a sufficient number of neutrons 
need to be available at the end of the $\nu p$-process.
In any case, relative to the solar pattern of Mo (Ru) isotopes, the 
production of $^{96}\mathrm{Mo}$ ($^{100}\mathrm{Ru}$) is always 
much less significant than that of $^{92,94}\mathrm{Mo}$ 
($^{96,98}\mathrm{Ru}$) for the conditions explored here
(see e.g., Fig.~\ref{fig:prod_fac}).

\subsection{Solar abundances of $^{92,94}$Mo and $^{96,98}$Ru}
\label{sec:p_isotopes}

\begin{figure*}[!hb]
\centering
    \includegraphics[width=0.43\textwidth]{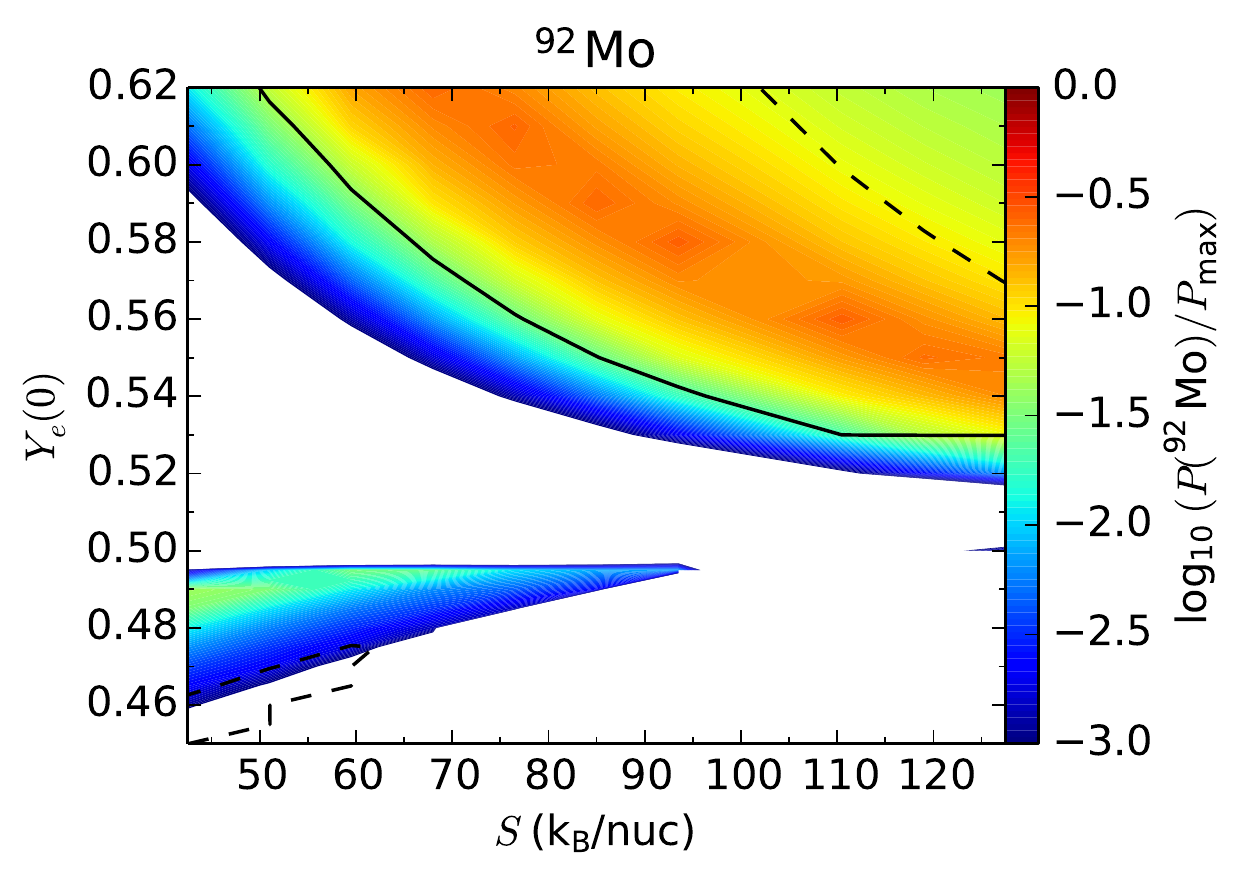}%
    \includegraphics[width=0.43\textwidth]{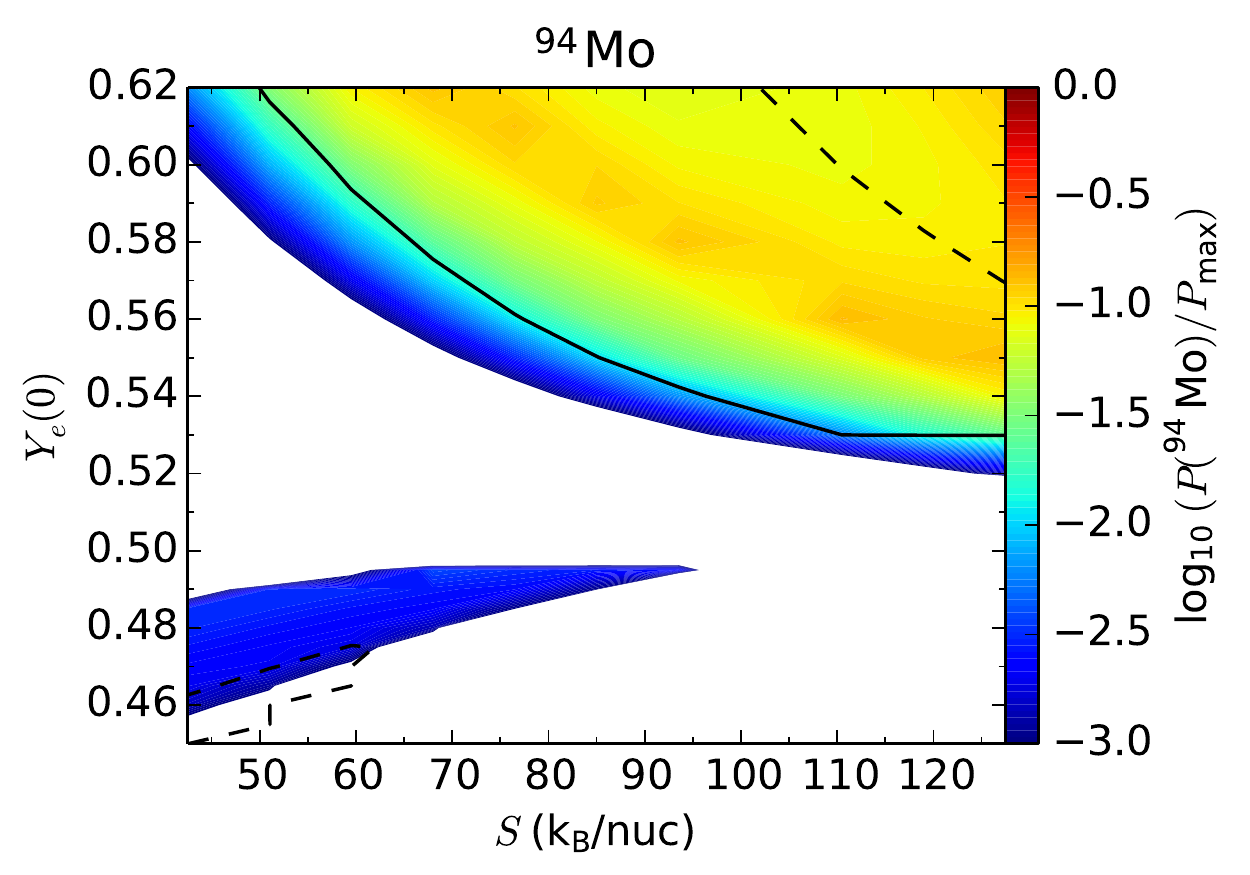}\\
    \includegraphics[width=0.43\textwidth]{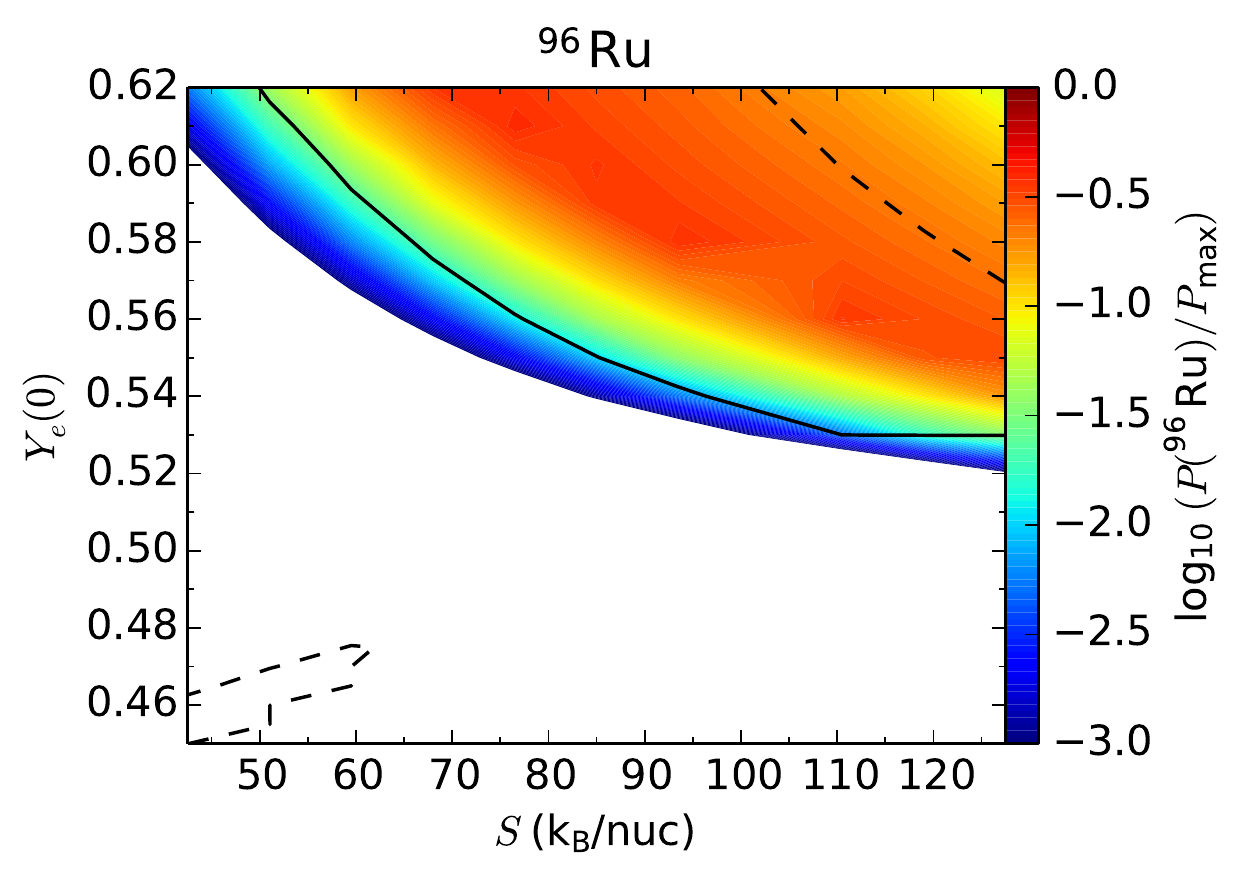}%
    \includegraphics[width=0.43\textwidth]{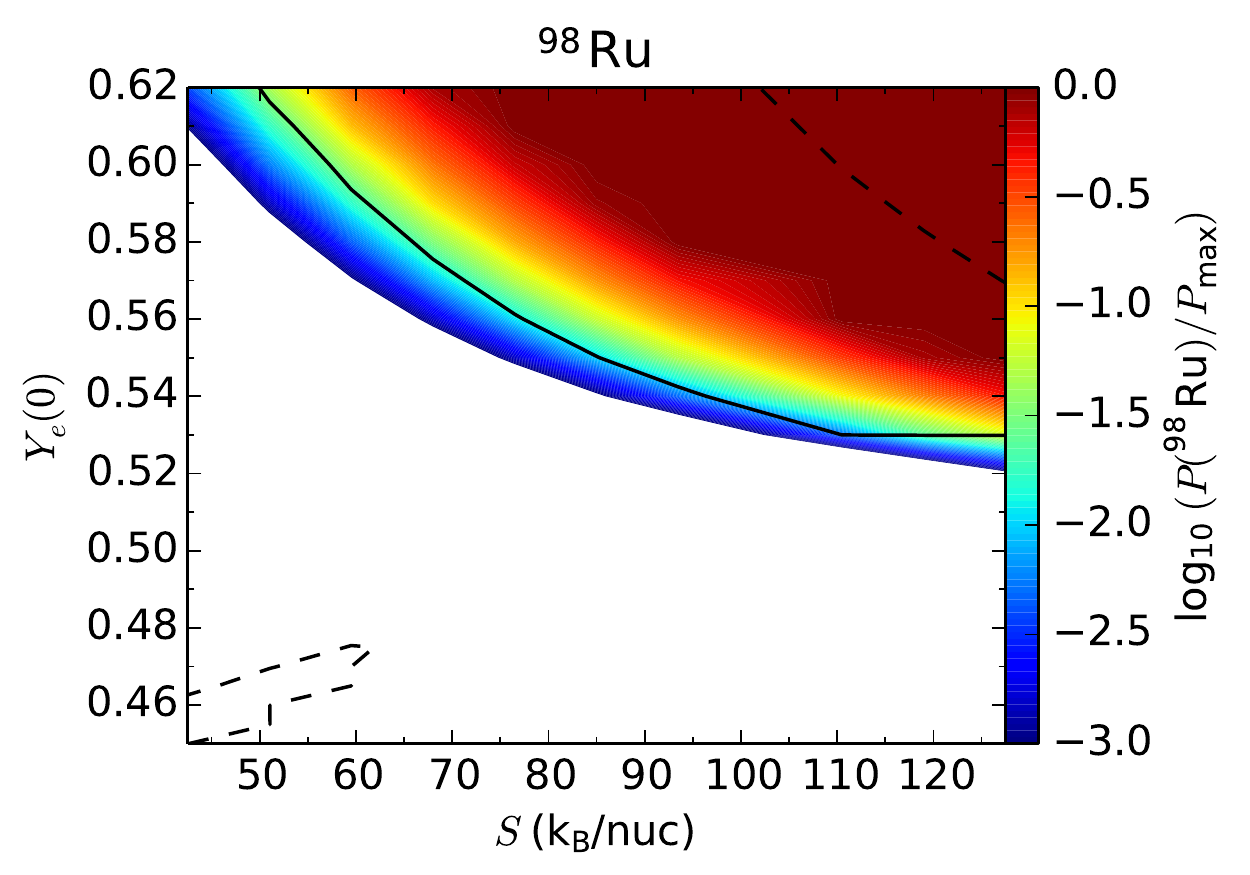}
    \caption{Color-coded contours illustrate the ratios (on a logarithmic scale) of the production factors of $^{92,94}$Mo (upper panels) 
    and $^{96,98}$Ru (lower panels) relative to the maximum production factor among all isotopes produced by the trajectory ejected
    at $t_{\rm pb}=8$~s for different $Y_{e}(0)$ and $S$. Ratios smaller than $10^{-3}$ are shown in
      white. For reference, the solid (dashed) curves
      correspond to conditions where the ratio $X(^{96}$Ru)/$X(^{98}$Ru)
      [$X(^{92}$Mo)/$X(^{94}$Mo)] produced by the trajectory reaches the solar system value with
      the abundances of $^{96,98}$Ru ($^{92,94}$Mo) exceeding $10^{-10}$.}
\label{fig:ab_ProdFac}
\end{figure*}

\begin{figure*}[!ht]
    \centering
\includegraphics[width=0.43\textwidth]{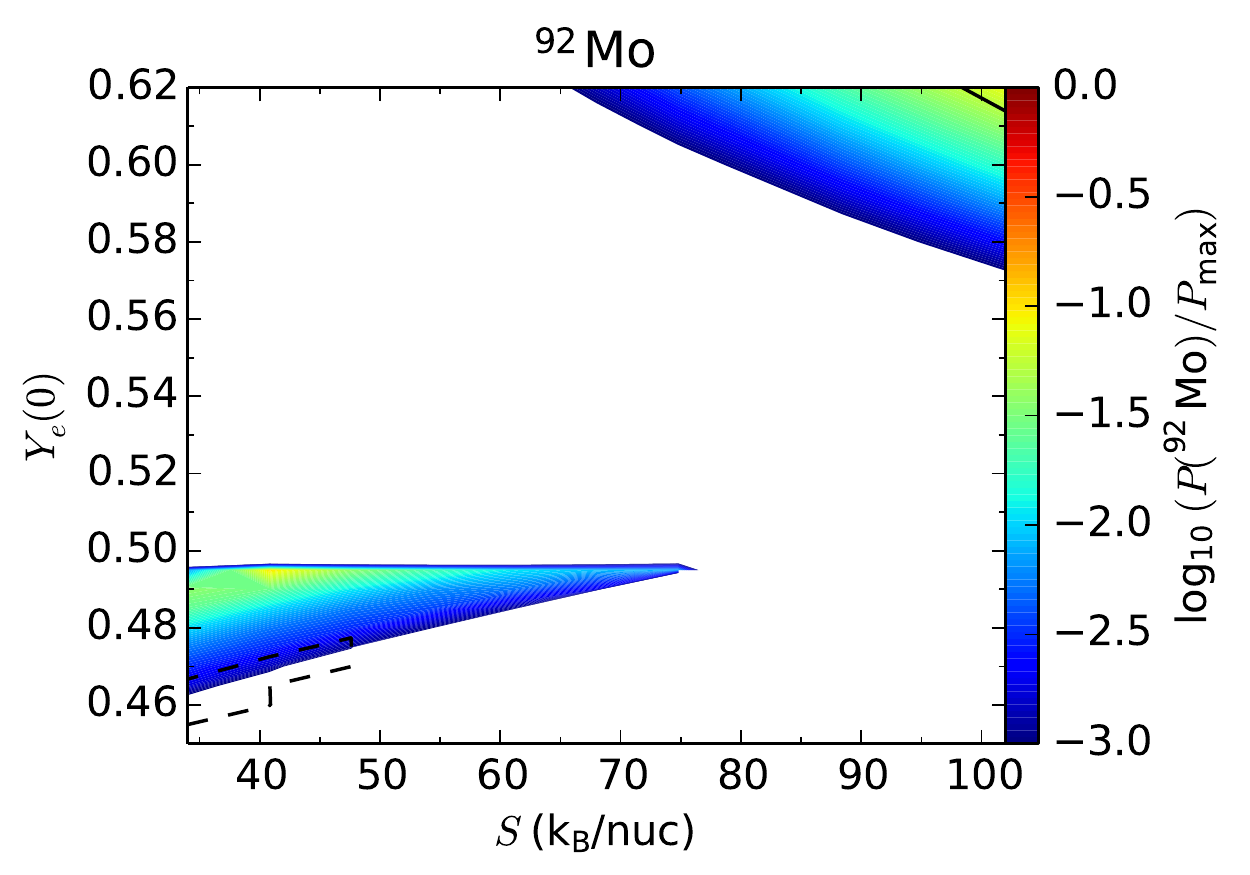}%
\includegraphics[width=0.43\textwidth]{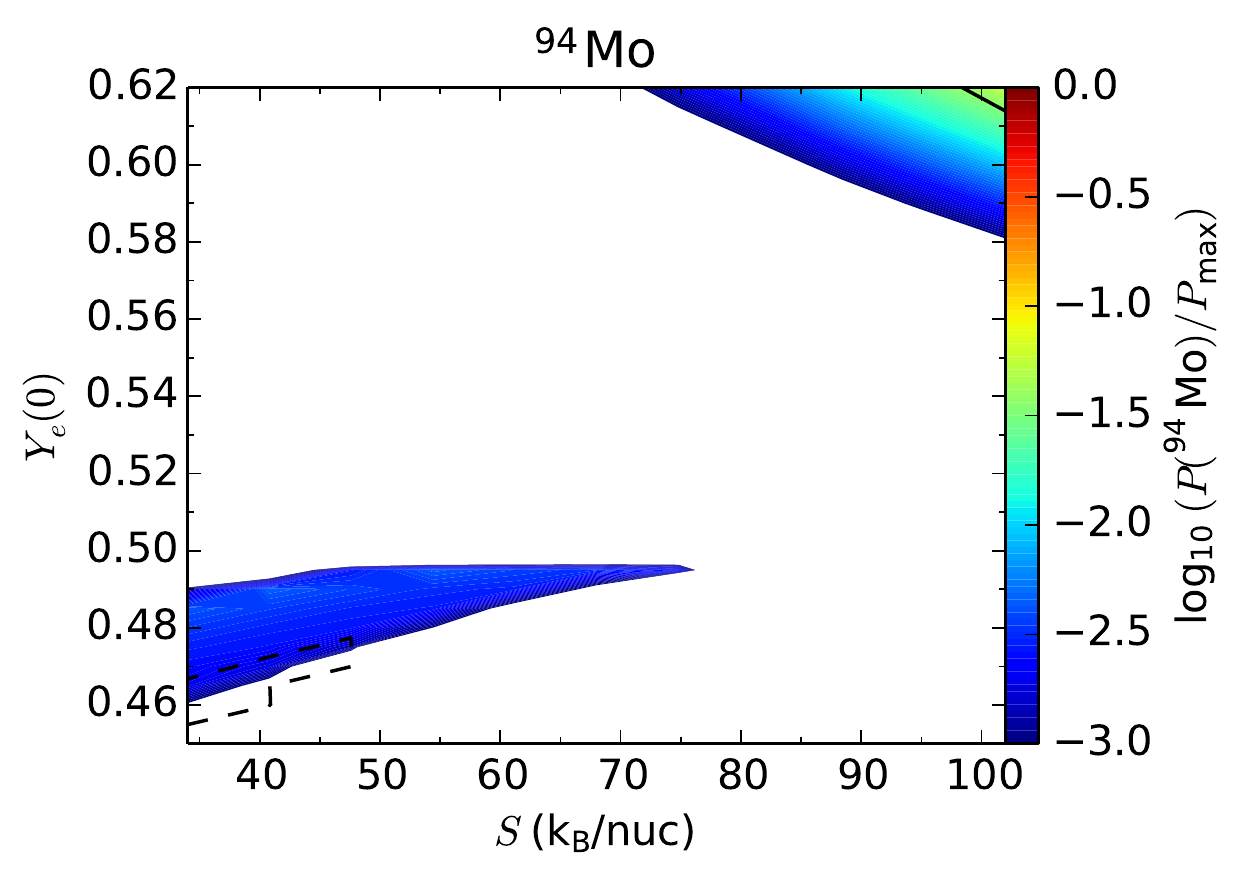}\\
\includegraphics[width=0.43\textwidth]{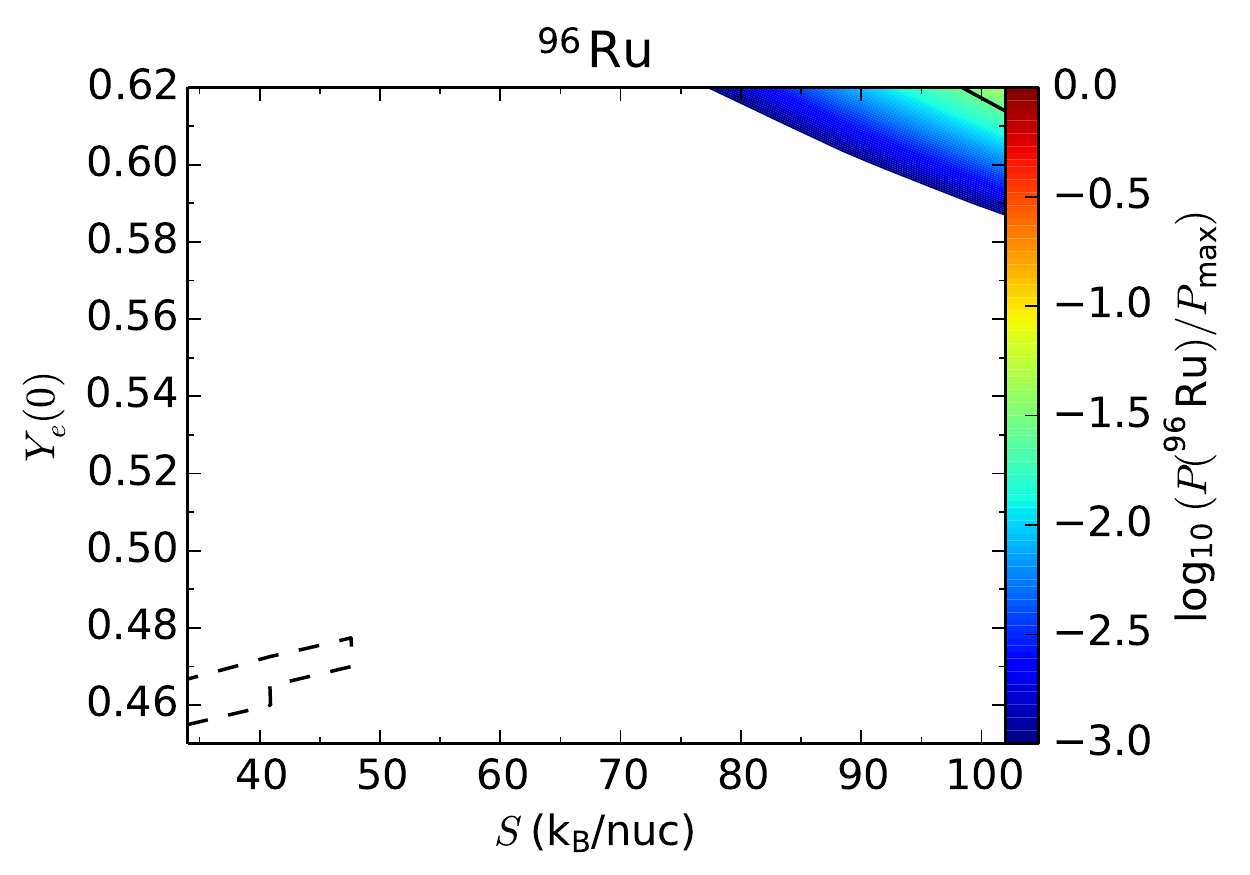}%
\includegraphics[width=0.43\textwidth]{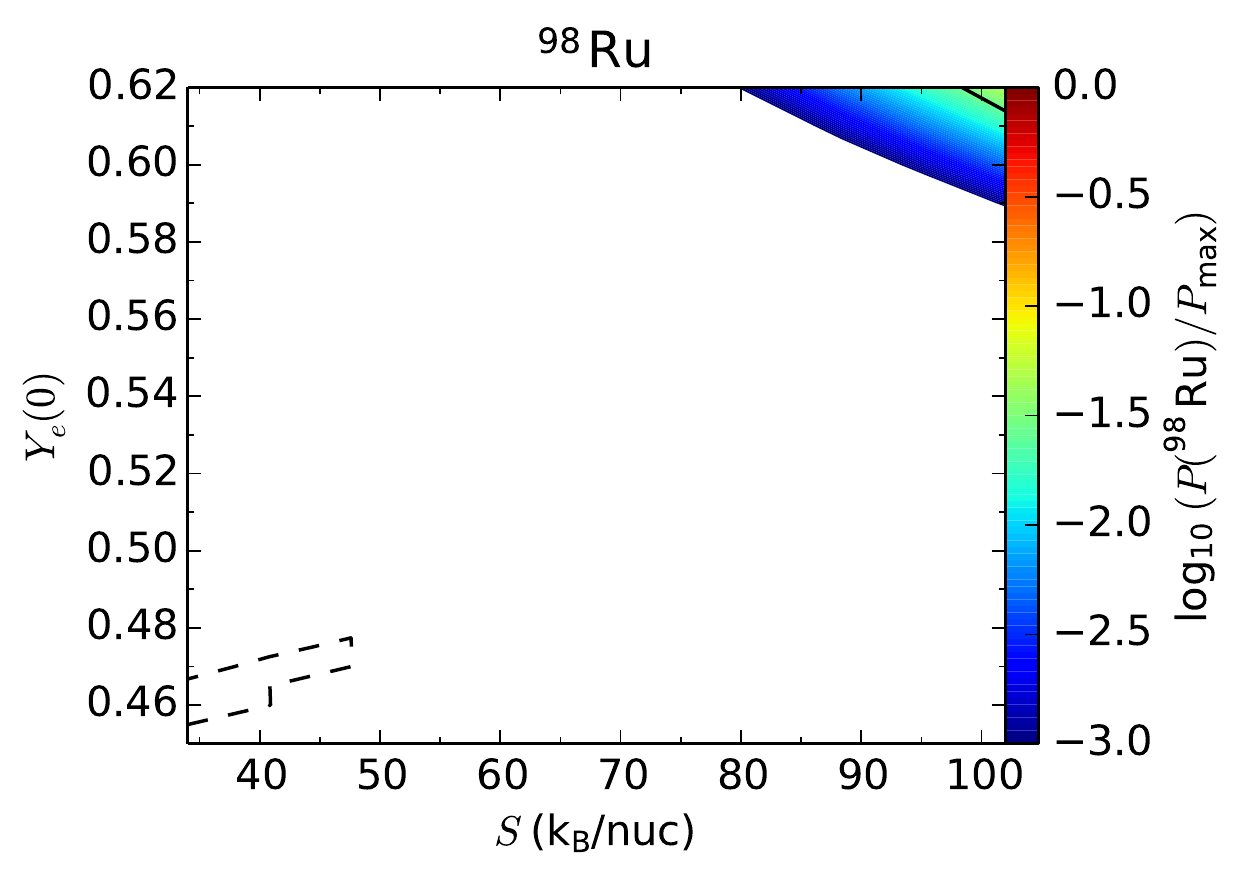}
\caption{Same as Fig.~\ref{fig:ab_ProdFac}, but for the trajectory ejected at $t_{\rm pb}=2$~s.}
\label{fig:ab_Mo_traj2s}
\end{figure*} 

\begin{figure*}[!htb]
\centering
\includegraphics[width=0.43\textwidth]{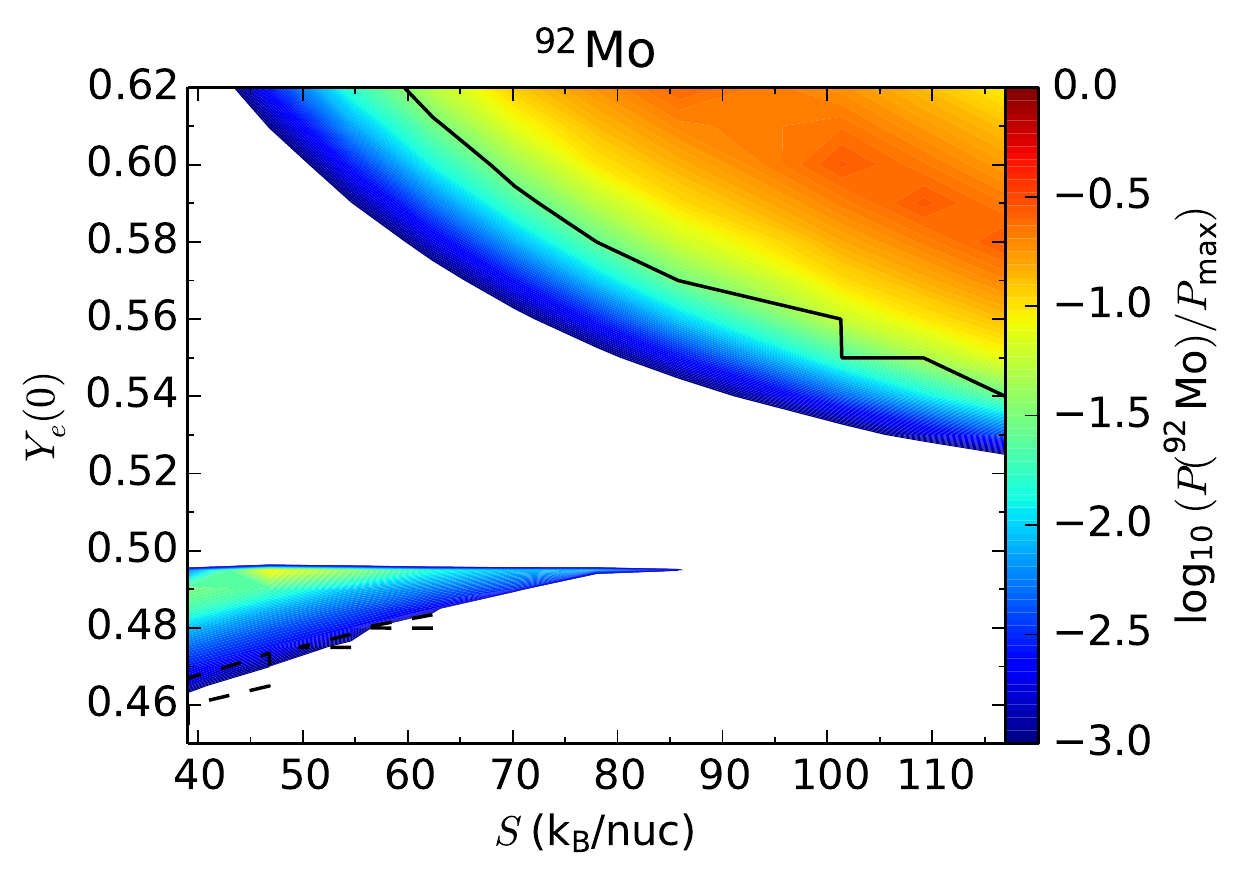}%
\includegraphics[width=0.43\textwidth]{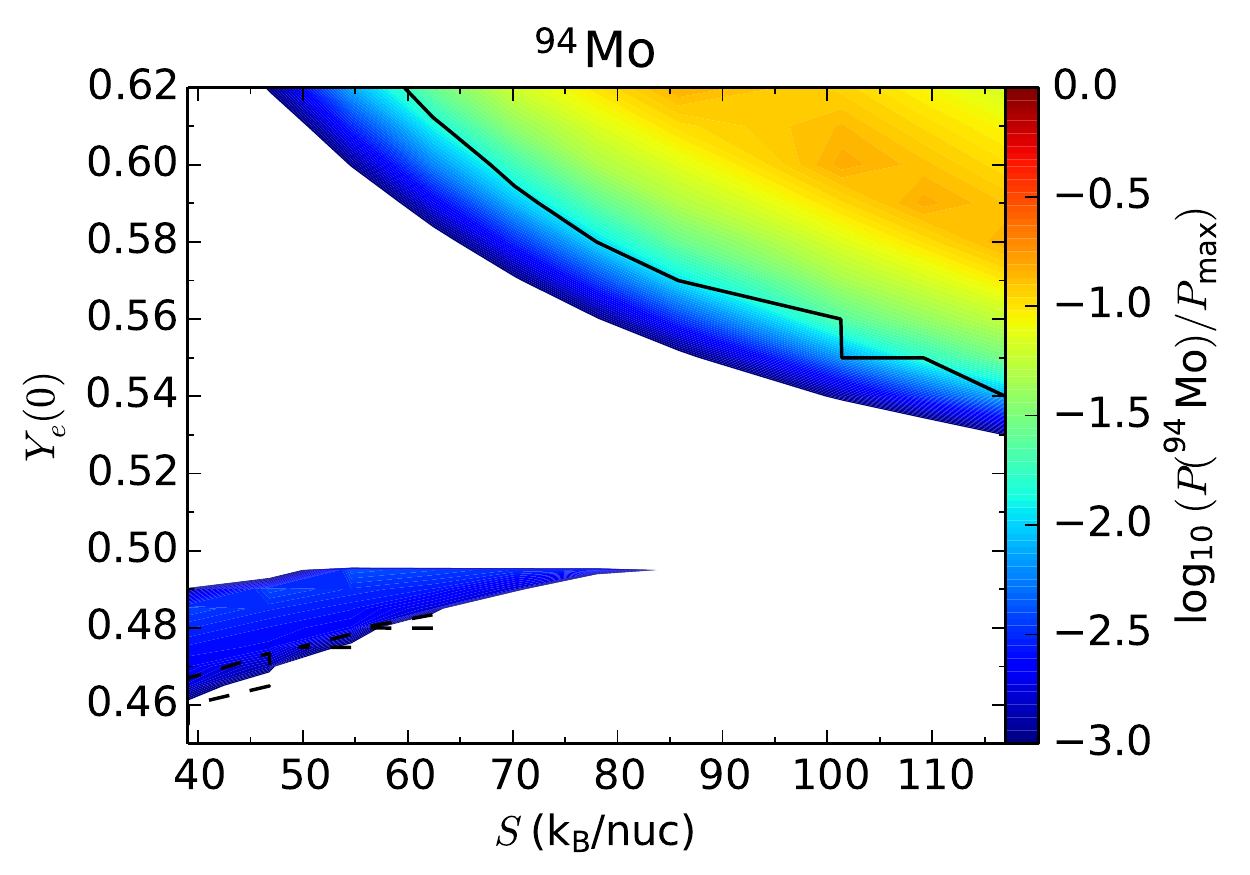}\\
\includegraphics[width=0.43\textwidth]{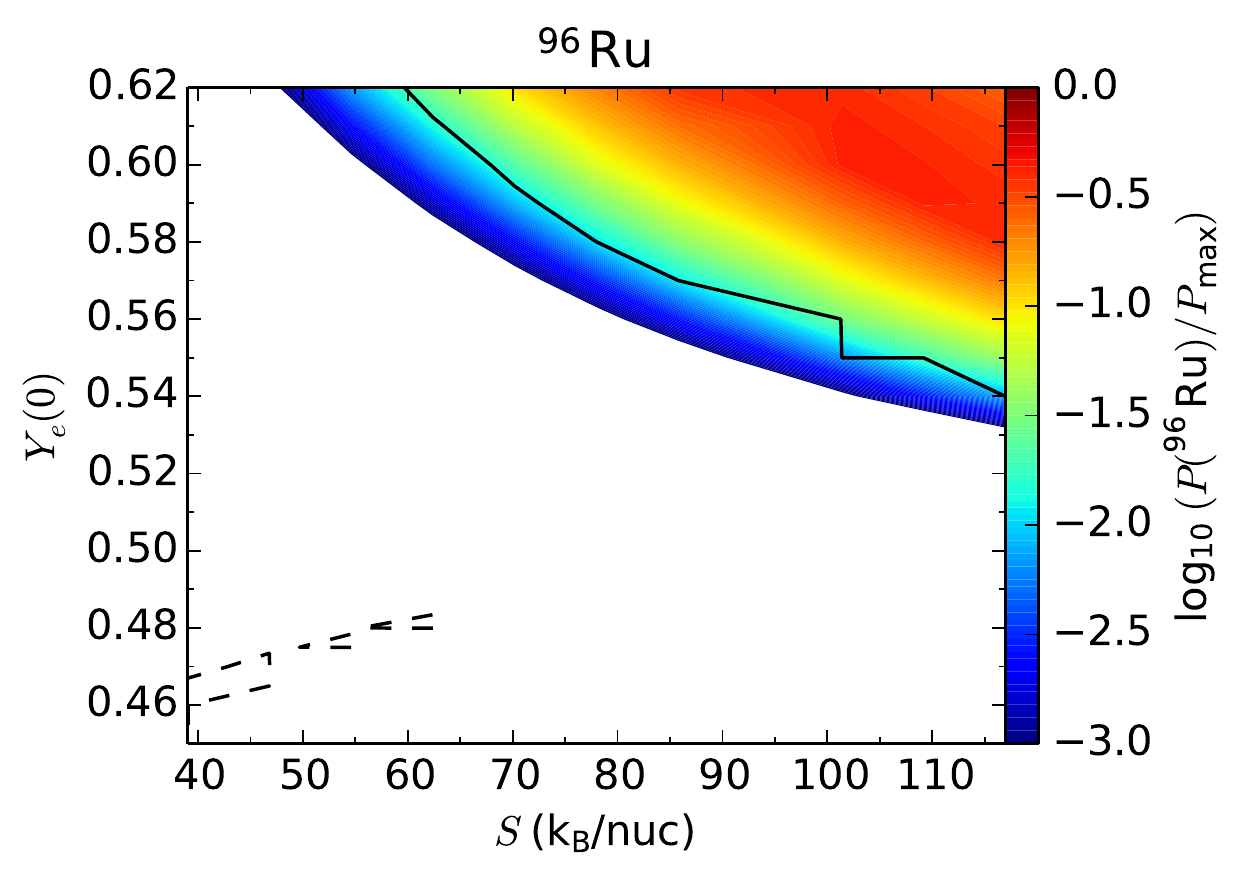}%
\includegraphics[width=0.43\textwidth]{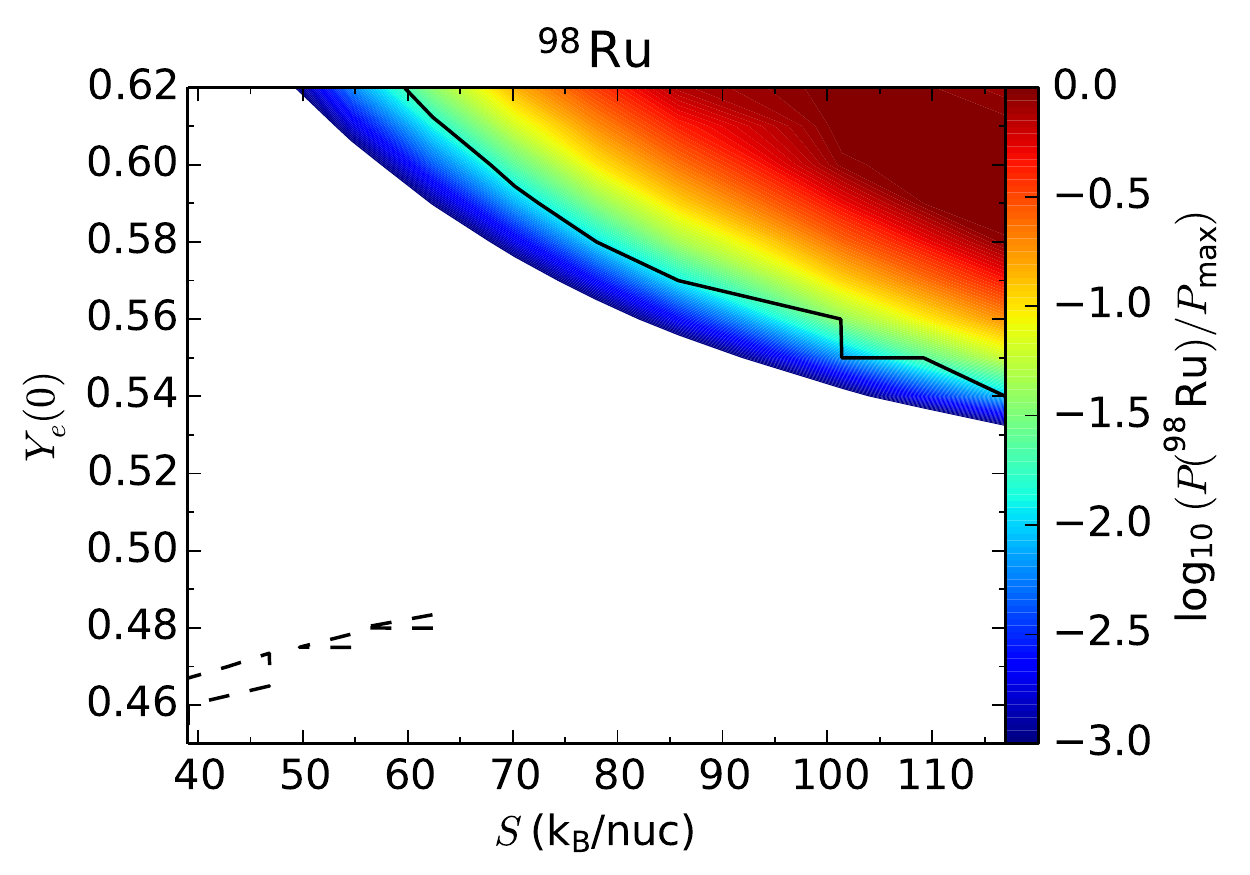}
\caption{Same as Fig.~\ref{fig:ab_ProdFac}, but for the trajectory ejected at $t_{\rm pb}=5$~s.}
\label{fig:ab_Mo_traj5s}
\end{figure*}   

In order for an astrophysical environment to be a major source
for the solar abundance of an isotope $^i$E, a necessary condition is that
this isotope must have a production factor $P(^i{\rm E})$ close to the maximum production factor $P_{\rm max}$
among all isotopes made in the same environment. The production factor $P(^i{\rm E})$ is defined as 

\begin{equation}
P(^i{\rm E})\equiv X(^i{\rm E})/X_\odot(^i{\rm E}),
\end{equation}
where $X(^i{\rm E})$ and $X_\odot(^i{\rm E})$ are the mass fraction of the isotope $^i$E produced in a model and observed in the solar system, respectively. The ratios $P(^i{\rm E})/P_{\rm max}$ for $^{92,94}$Mo and $^{96,98}$Ru produced by the trajectory ejected at
$t_{\rm pb}=8$~s are shown as color-coded contours for different $S$ and $Y_e(0)$ in Fig.~\ref{fig:ab_ProdFac}.

It can be seen from Fig.~\ref{fig:ab_ProdFac} that only $^{98}$Ru can be produced 
with the maximum production factor in the wind. In fact, while $P(^{98}{\rm Ru})/P_{\rm max}\sim 1$
is reached at $S\sim 128$~$k_{\mathrm{B}}$/nuc and $Y_e(0)\sim 0.61$, major production of $^{98}$Ru
occurs for a substantial range of conditions, e.g., $S\sim 80$--128~$k_{\mathrm{B}}$/nuc for $Y_e(0)\sim 0.6$
or $Y_e(0)\sim0.56$--0.62 for $S\sim 110$~$k_{\mathrm{B}}$/nuc. In contrast, the highest value of $P(^i{\rm E})/P_{\rm max}$
is $\sim 0.4$ [at $S\sim 77$~$k_{\mathrm{B}}$/nuc, $Y_e(0)\sim 0.61$] for $^{96}$Ru, $\sim 0.27$ [at $S\sim 111$~$k_{\mathrm{B}}$/nuc, $Y_e(0)\sim 0.56$] for $^{92}$Mo,
and $\sim 0.14$ [at $S\sim 111$~$k_{\mathrm{B}}$/nuc, $Y_e(0)\sim 0.56$] for $^{94}$Mo. These results suggest that
proton-rich winds can make dominant contributions to the solar abundance of $^{98}\mathrm{Ru}$, 
significant contributions to those of $^{96}$Ru ($\lesssim 40\%$) and $^{92}$Mo 
($\lesssim 27\%$), and relatively minor contributions to that of $^{94}$Mo 
($\lesssim 14\%$). Figure~\ref{fig:ab_ProdFac} also shows that neutron-rich winds make negligible 
contributions to the solar abundances of $^{92,94}$Mo and cannot produce any significant amounts of $^{96,98}$Ru.

The above results strongly suggest that sources other than the neutrino-driven wind, e.g., 
Type Ia supernovae (see \citealt{Travaglio.etal:2014} for a recent study), are required to account for the
solar abundances of $^{92,94}$Mo and $^{96}$Ru. Results for the trajectories ejected at $t_{\rm pb}=2$ and 5~s
also support this conclusion (see Figs.~\ref{fig:ab_Mo_traj2s} and \ref{fig:ab_Mo_traj5s}).
While proton-rich winds can make dominant contributions to the solar abundance of $^{98}\mathrm{Ru}$,
the exact contribution from this source can be determined only when contributions from other 
sources are established.

We emphasize that in considering potential contributions from a source to the solar 
abundances of $^{92,94}$Mo and $^{96,98}$Ru, the associated $P(^i{\rm E})/P_{\rm max}$ values are a
critical test. The production ratio of $^{92,94}$Mo ($^{96,98}$Ru) relative to the
solar value is secondary in that it is important only when the production factors for both isotopes 
are close to $P_{\rm max}$. Because only $^{98}$Ru can have $P(^{98}{\rm Ru})/P_{\rm max}\sim 1$
in the wind, explanation of the ratios ($^{92}$Mo/$^{94}$Mo$)_\odot=1.60$ and 
($^{96}$Ru/$^{98}$Ru$)_\odot=2.97$ \citep{Lodders:2003} for the solar system crucially depends on 
other sources for these isotopes. In this regard, although both these ratios can be achieved in the 
wind (see Figs.~\ref{fig:ab_ProdFac}--\ref{fig:ab_Mo_traj5s}), this result is largely irrelevant for
explaining the relative abundances of $^{92,94}$Mo ($^{96,98}$Ru) in the solar system.

\begin{figure}[!ht]
  \centering
  \includegraphics[width=0.85\linewidth]{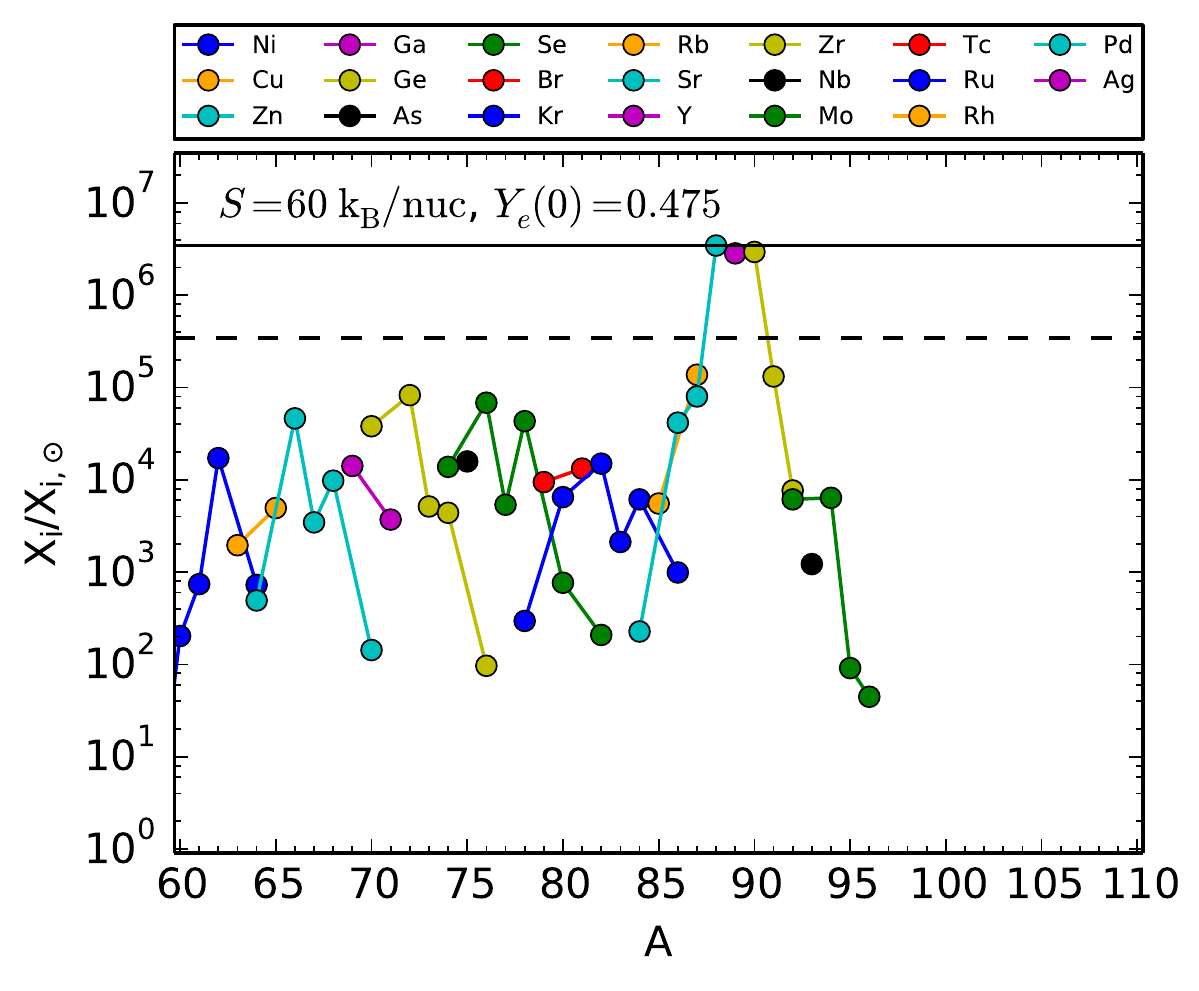}\\
  \includegraphics[width=0.85\linewidth]{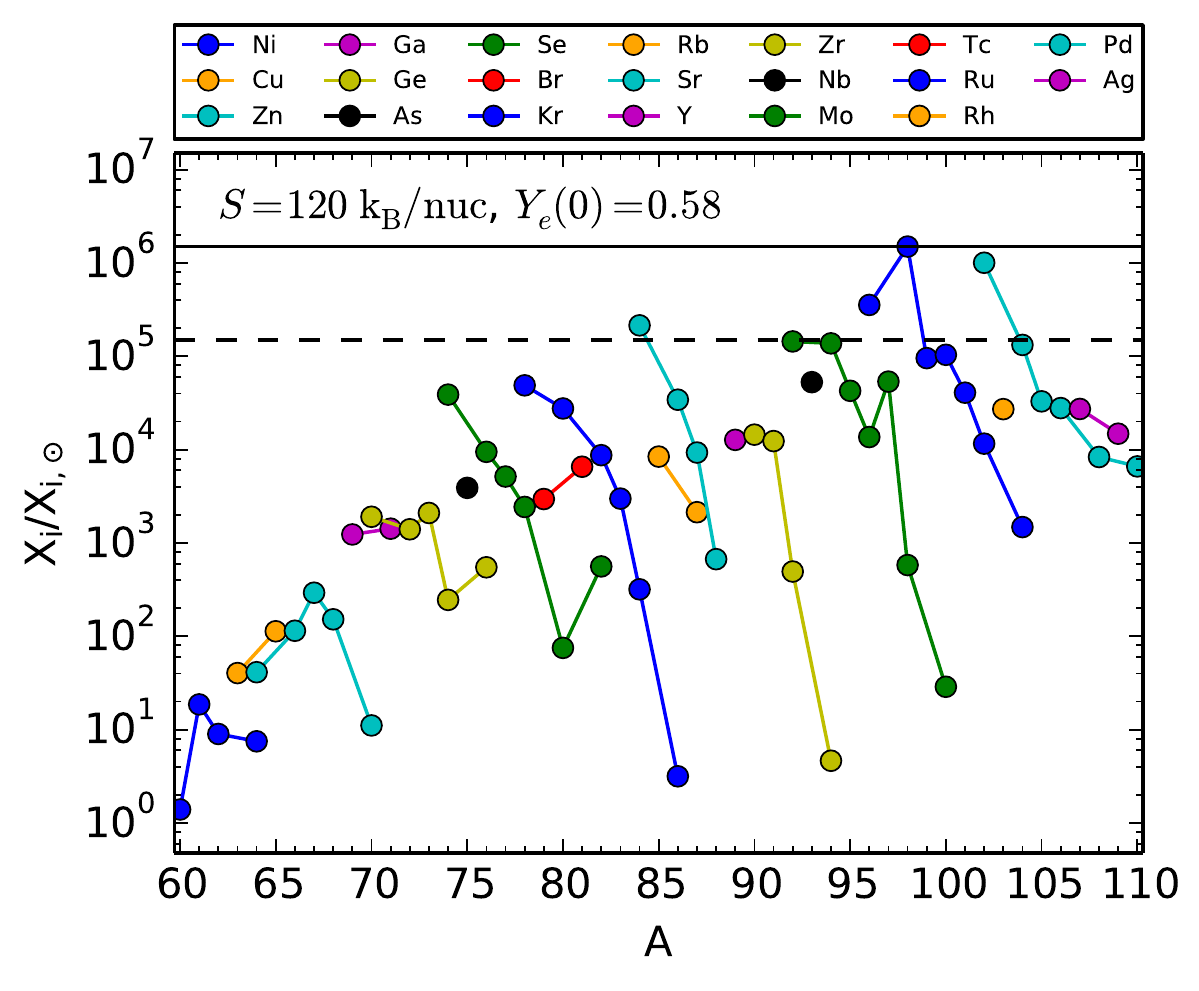}\\
  \includegraphics[width=0.85\linewidth]{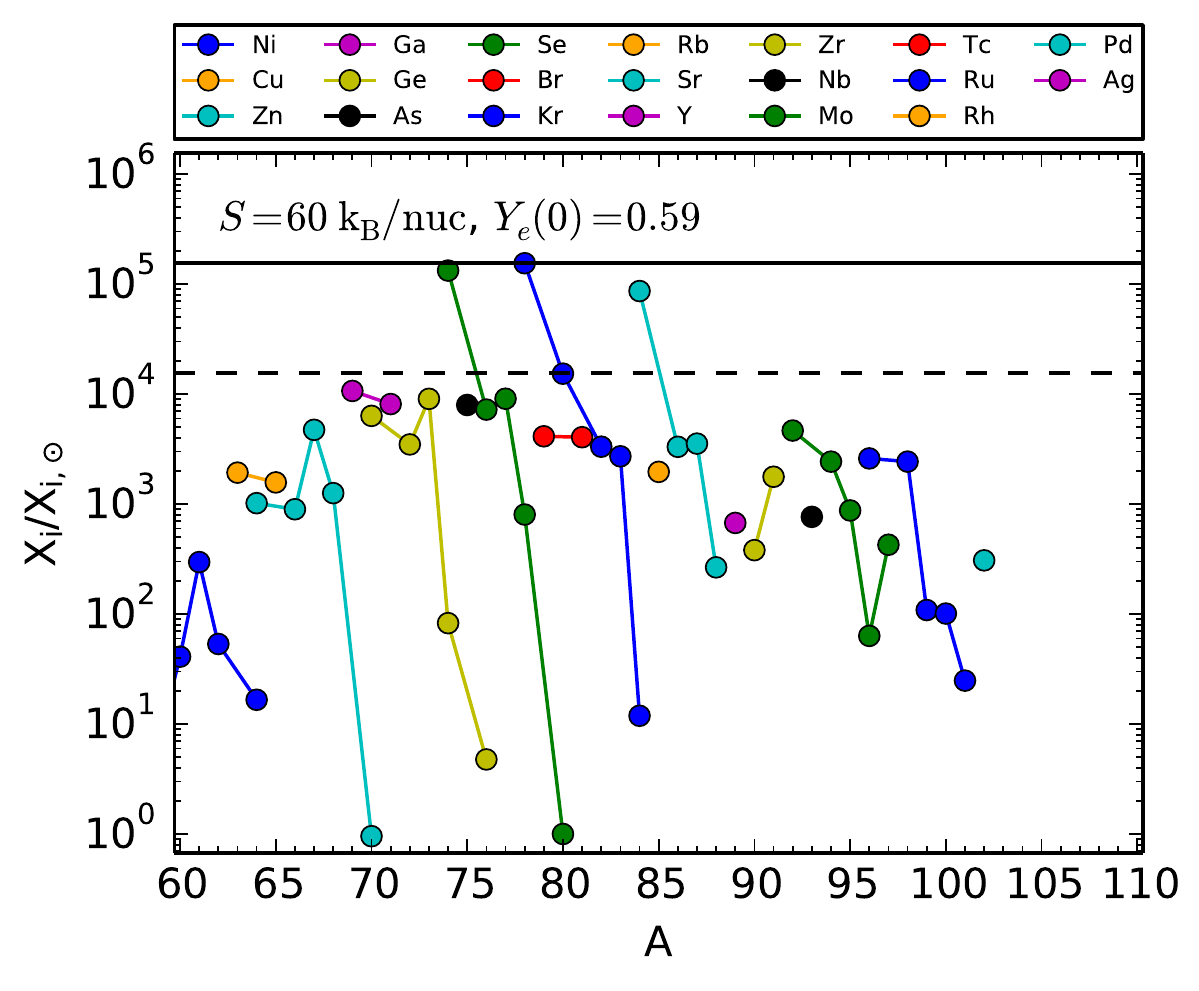}
  \caption{The ratio of the mass fraction of an isotope produced by the wind trajectory ejected at
  $t_{\rm pb}=8$~s ($X_i$) relative to that
  observed in the solar system ($X_{i,\odot}$), i.e., the production factor, as a function of the mass 
  number ($A$). Isotopes from the same element are connected by line segments. 
  The horizontal lines indicate a normalization band given by the largest production 
  factor (solid line) and a factor 10 less (dashed line). Nuclei falling within the band are 
  the main products. The solar ratio of $^{92,94}$Mo is achieved in the top and middle panels,
  while that of $^{96,98}$Ru is achieved in the bottom panel. However, the actual ratio of $^{92,94}$Mo 
  in the solar system must be explained mostly by sources other than the wind, and that of $^{96,98}$Ru 
  by combining the wind and other sources.}
  \label{fig:prod_fac}
  \end{figure}

For illustration, we show in Fig.~\ref{fig:prod_fac} models based on the trajectory
ejected at $t_{\rm pb}=8$~s that produce $^{92,94}$Mo ($^{96,98}$Ru) in the solar ratio but have 
little to do with accounting for their abundance ratio in the solar system. The solar ratio for $^{92,94}$Mo 
can be achieved in neutron-rich 
winds for $S=60$~$k_{\mathrm{B}}$/nuc and $Y_e (0)= 0.475$ (top panel) or in proton-rich winds for $S=120$~$k_{\mathrm{B}}$/nuc and 
$Y_e (0)= 0.58$ (middle panel). However, in neutron-rich winds, the predominantly 
produced isotopes are $^{88}$Sr, $^{89}$Y, and $^{90}$Zr with the magic neutron number
$N=50$ \citep{Hoffman.etal:1997, Hoffman.etal:1996, Witti.etal:1994}, while
the production factors for $^{92,94}$Mo are only $\approx2 \times10^{-3}P_{\mathrm{max}}$.
These production factors dramatically increase to $\approx0.1P_{\mathrm{max}}$ for 
proton-rich winds, but in this case $^{98}$Ru has the largest production factor and 
$^{96}$Ru is coproduced in a non-solar ratio. Nevertheless, this case represents approximately 
the optimal scenario for wind contributions to the solar abundances of $^{92,94}$Mo and $^{96,98}$Ru. 
Finally, the solar ratio for $^{96,98}$Ru can be achieved in proton-rich winds for $S=60$~$k_{\mathrm{B}}$/nuc and 
$Y_e (0)= 0.59$ (bottom panel). However, for these conditions the predominantly produced 
isotopes are $^{74}$Se, $^{78}$Kr, and $^{84}$Sr.

\cite{Fisker.et.al.2009} found that $^{92,94}$Mo could be produced in the solar ratio 
in slightly proton-rich winds but with too small production factors to account for their solar abundances.
Those results are qualitatively consistent with ours, although \cite{Fisker.et.al.2009} used very different 
trajectories.

\subsection{Peculiar {\rm Mo} patterns in {\rm SiC X} grains}
\label{sec:9597Mo}  

Relative to $^{96}$Mo, an $s$-only isotope, the $p$-only isotopes
$^{92,94}$Mo and the $r$-only isotope $^{100}$Mo are nearly
absent in mainstream SiC grains. This points to an $s$-process origin
in AGB stars for the Mo patterns in such grains
(see e.g., \citealt{Lugaro.etal:2003} for a detailed study).
In contrast, SiC X grains have peculiar Mo patterns with large enrichments in 
the mixed isotopes $^{95,97}$Mo \citep{Pellin.etal:1999,Pellin.etal:2006}. 
Some of the X grains are also highly enriched in the $r$-only isotope $^{96}$Zr 
\citep{Davis.etal:1999,Pellin.etal:2006}.
While an $r$-process origin may possibly account for the large enrichments in 
$^{96}$Zr and $^{95,97}$Mo, this is inconsistent with the data on
the $r$-only isotope $^{100}$Mo, which is not significantly 
enriched in most SiC X grains \citep{Pellin.etal:1999,Pellin.etal:2006}.

\cite{Meyer:2000} proposed to explain the overabundance of $^{95,97}\mathrm{Mo}$ 
in SiC X grains with a neutron burst model \citep{Howard.etal:1992}. They first
exposed a solar distribution of nuclei to a weak neutron fluence to mimic the weak 
$s$-process during the presupernova phase of a massive star
(see e.g.,~\citealt{Pignatari.etal:2010,Kaeppeler.etal:2011}). Then they abruptly
heated the processed matter to 1~GK to mimic the effect of a supernova shock
and allowed the shocked matter to expand and cool. The burst of neutrons released 
by $(\alpha,n)$ reactions redistributed the initial 
abundances of Y and Zr isotopes to heavier isotopes up to $A\sim97$, but the
burst was not strong enough to accumulate much matter at $^{100}$Zr. The original
abundances of Mo isotopes were also redistributed to heavier isotopes. Finally,
large abundances of $^{95,97}$Mo were obtained upon the $\beta$ decay of
$^{95}$Y and $^{95,97}$Zr that were newly synthesized by the neutron burst.

\begin{figure}[]
  \centering
  \includegraphics[width=0.85\linewidth]{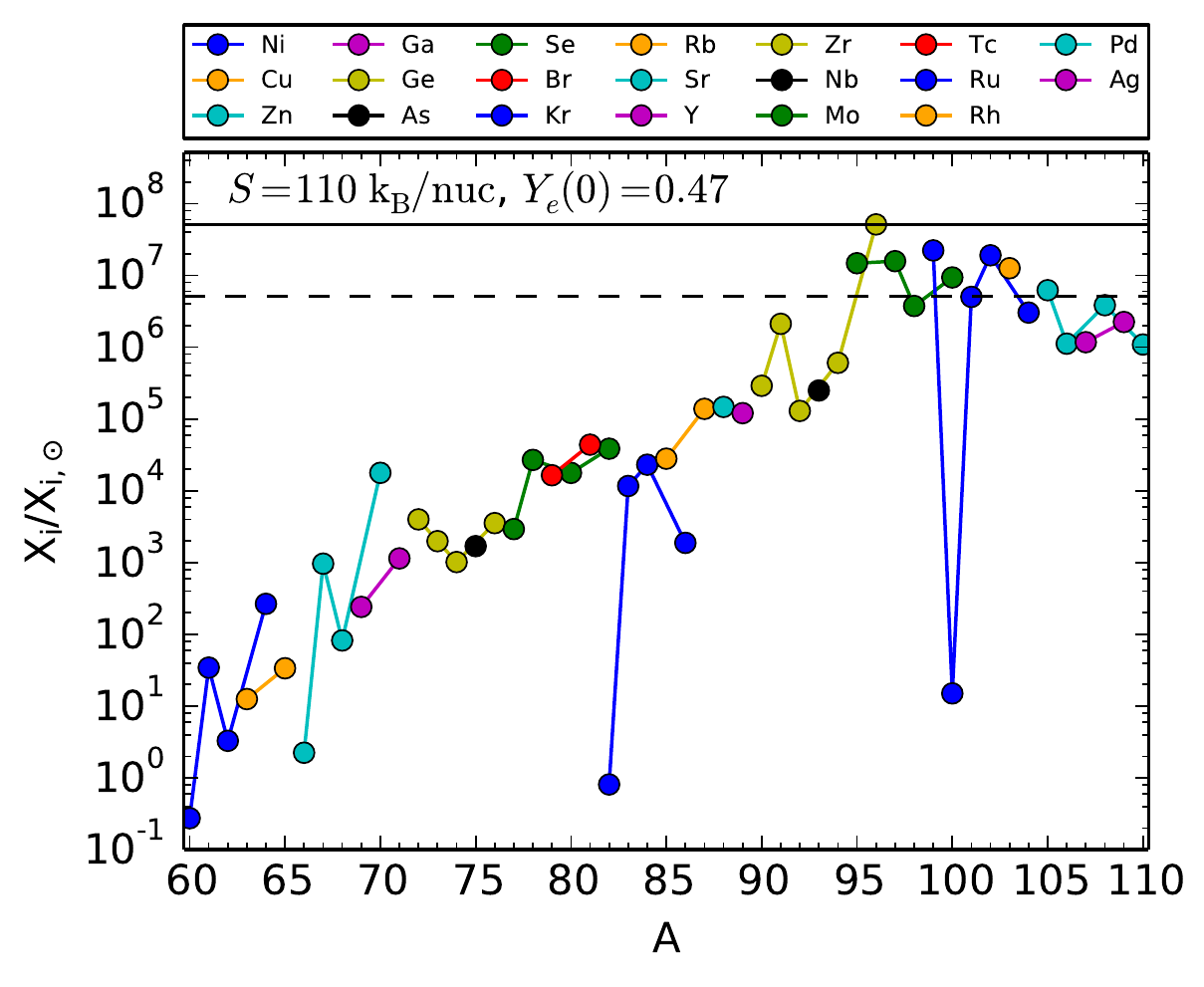}
  \caption{Same as Fig.~\ref{fig:prod_fac}, but for wind conditions that are relevant
  for explaining the peculiar patterns of
  $^{95,97,98,100}\mathrm{Mo}$ found in SiC X grains.}
    \label{fig:prod_sicx}
\end{figure}

While the above neutron burst model offers a potential explanation of the enrichments
in $^{96}$Zr and $^{95,97}$Mo in SiC X grains, it remains to be seen if the conditions 
assumed could be provided by a detailed astrophysical model. Here we explore
another potential explanation based on the neutrino-driven wind. 
Figure~\ref{fig:prod_sicx} shows the production factors for various isotopes made
in the wind that corresponds to the trajectory ejected at $t_{\rm pb}=8$~s with 
$S=110$~$k_{\mathrm{B}}$/nuc and $Y_e(0)=0.47$. It can be seen that $^{96}\mathrm{Zr}$ has the largest 
production factor and $^{95,97,98,100}$Mo are also significantly produced.
Further, the production factors for $^{95,97}$Mo exceed those for $^{98,100}$Mo,
in agreement with the patterns in SiC X grains \citep{Pellin.etal:1999,Pellin.etal:2006}.
Specifically, the ratios of the production factors are
$P(^{95}{\rm Mo}):P(^{97}{\rm Mo}):P(^{98}{\rm Mo}):P(^{100}{\rm Mo})=
1.5:1.64:0.393:1$. Note also that neither the $p$-only isotopes $^{92,94}$Mo nor
the $s$-only isotope $^{96}$Mo is produced in this wind.

Because the sources for $^{92,94}$Mo are uncertain, we focus on explaining
the patterns of $^{95,96,97,98,100}$Mo in SiC X grains. We consider that these five
isotopes can be accounted for by mixtures of contributions from the $s$-process,
the $r$-process, and the neutrino-driven wind. 
Specifically, we assign the $s$-only isotope $^{96}$Mo
exclusively to the $s$-process and use its abundance in a grain along with the
solar $s$-process pattern to determine the $s$-process contributions to other Mo isotopes.  
Because the ``$r$-only'' isotope $^{100}$Mo cannot be made in the $s$-process,
we assign a fraction $f_w$ of its abundance in a grain to the wind and the rest to
the $r$-process. The $r$-process fraction $(1-f_w)$ of the $^{100}$Mo abundance 
in a grain is used along with the solar $r$-process pattern to determine the $r$-process 
contributions to other Mo isotopes. Consequently, the abundance of the isotope $^i$Mo 
in a grain, $(^i{\rm Mo})_g$, is given by
\begin{eqnarray}
\label{eq:imo}
&&(^i{\rm Mo})_g=\left(\frac{^i{\rm Mo}}{^{96}{\rm Mo}}\right)_s(^{96}{\rm Mo})_g+\\
&&\left[(1-f_w)\left(\frac{^i{\rm Mo}}{^{100}{\rm Mo}}\right)_r+
f_w\left(\frac{^i{\rm Mo}}{^{100}{\rm Mo}}\right)_w\right](^{100}{\rm Mo})_g,\nonumber
\end{eqnarray}
where the $s$-process, $r$-process, and wind production ratios 
$(^i{\rm Mo}/^{96}{\rm Mo})_s$, $(^i{\rm Mo}/^{100}{\rm Mo})_r$, and
$(^i{\rm Mo}/^{100}{\rm Mo})_w$ are assumed to be fixed. 

Equation~(\ref{eq:imo}) can be rewritten as
\begin{eqnarray}
\label{eq:imo2}
&&\frac{(^i{\rm Mo}/^{96}{\rm Mo})_g}{(^i{\rm Mo}/^{96}{\rm Mo})_\odot}
=f_{s,\odot}(^i{\rm Mo})+\\
&&\left[(1-f_w)f_{r,\odot}(^i{\rm Mo})+f_w\frac{P(^i{\rm Mo})}{P(^{100}{\rm Mo})}\right]
\frac{(^{100}{\rm Mo}/^{96}{\rm Mo})_g}{(^{100}{\rm Mo}/^{96}{\rm Mo})_\odot},\nonumber
\end{eqnarray}
where $f_{s,\odot}(^i{\rm Mo})$ and $f_{r,\odot}(^i{\rm Mo})=1-f_{s,\odot}(^i{\rm Mo})$
are the $s$-process and $r$-process fractions of the solar $^i$Mo abundance,
respectively. We take $f_{s,\odot}(^i{\rm Mo})=0.50$, 0.59, 0.75, and 0 for
$^{95}$Mo, $^{97}$Mo, $^{98}$Mo, and $^{100}$Mo, respectively.
These values are consistent with both the estimates of \cite{Arlandini:1999}
and the $s$-process patterns found in mainstream SiC grains
(e.g., \citealt{Pellin.etal:1999}). Note that 
$f_{r,\odot}(^i{\rm Mo})=1-f_{s,\odot}(^i{\rm Mo})$ is valid when the wind makes
negligible contributions to the solar abundances of the relevant Mo isotopes.

\begin{figure}[!hb]
  \centering
  \includegraphics[width=0.4\textwidth]{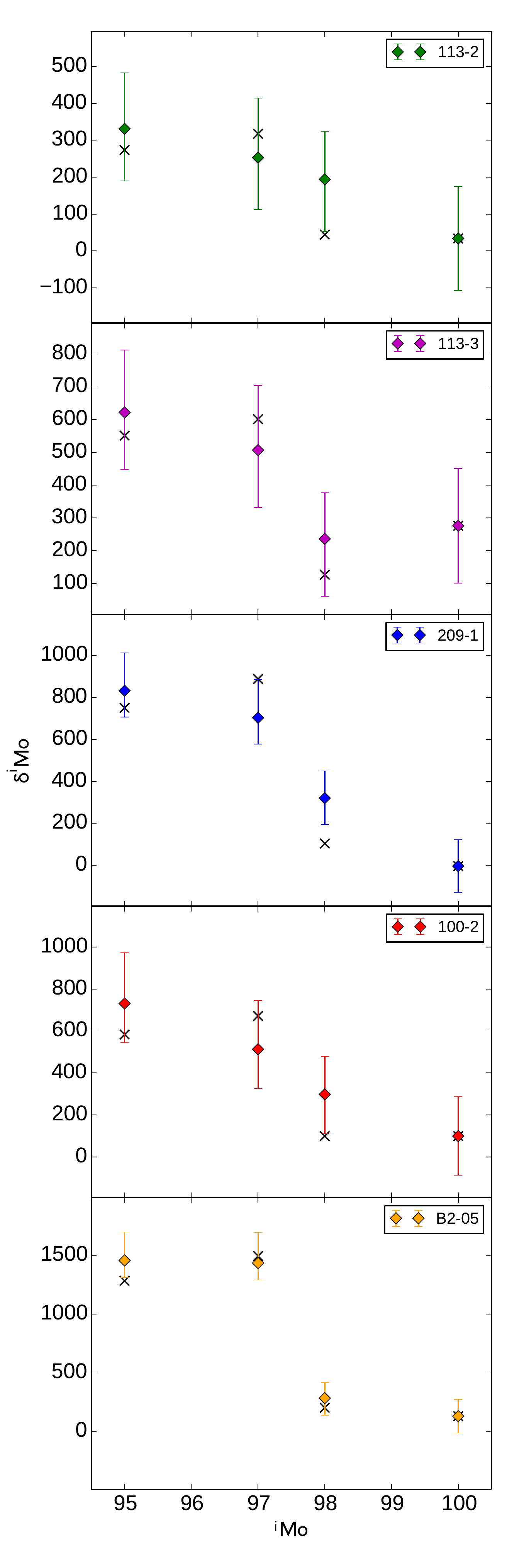}
  \caption{Comparison of the model (crosses) including wind contributions and the data
  (colored diamonds with error bars) on the peculiar patterns of $^{95,97,98,100}$Mo in 
  five SiC X grains \citep{Pellin.etal:2006}.}
    \label{fig:9597Mo}  
\end{figure}

In terms of the meteoritic notation
\begin{equation}
\delta\,^i{\rm Mo}\equiv 1000\times\left[\frac{(^i{\rm Mo}/^{96}{\rm Mo})_g}
{(^i{\rm Mo}/^{96}{\rm Mo})_\odot}-1\right],
\end{equation}
Eq.~(\ref{eq:imo2}) can be written as
\begin{eqnarray}
\label{eq:idel}
&&\delta\,^i{\rm Mo}=10^3f_w
\left[\frac{P(^i{\rm Mo})}{P(^{100}{\rm Mo})}-f_{r,\odot}(^i{\rm Mo})\right]+\\
&&\left[(1-f_w)f_{r,\odot}(^i{\rm Mo})+f_w\frac{P(^i{\rm Mo})}{P(^{100}{\rm Mo})}\right]
\delta\,^{100}{\rm Mo}.\nonumber
\end{eqnarray}
Using the central value of $\delta\,^{100}{\rm Mo}$ for a grain along with 
the wind production factor $P(^i{\rm Mo})$ relative to
$P(^{100}{\rm Mo})$ and the solar
$r$-process fraction $f_{r,\odot}(^i{\rm Mo})$ given above,
we find $f_w$ for which Eq.~(\ref{eq:idel}) gives $\delta\,^i{\rm Mo}$ in good 
agreement with the data on $^{95,97,98}$Mo in the same grain.
These results are shown in Fig.~\ref{fig:9597Mo} with
$f_w=0.23$, 0.3, 0.7, 0.45, and 1 for
five SiC X grains \mbox{113-2}, \mbox{113-3}, \mbox{209-1}, \mbox{100-2}, and \mbox{B2-05} \citep{Pellin.etal:2006},
respectively. It can be seen that the above simple model including the wind
contributions can explain the overall peculiar patterns of $^{95,97,98,100}$Mo 
found in these grains. We expect that better agreement with the data, e.g., 
elimination of the discrepancy for $^{98}$Mo in the grain 209-1, may be achieved
by varying $\delta\,^{100}{\rm Mo}$ within the measurement errors or using
different wind conditions to optimize the wind production factors, or both. However, 
we consider that the results shown in Fig.~\ref{fig:9597Mo} are sufficient as a 
proof of concept.

{\section{Conclusions}
\label{sec:summary}
We have presented a detailed parametric study on the production 
of Mo and Ru isotopes in neutrino-driven winds. Our results can be generalized 
if conditions similar to those studied here are also obtained for other types of 
ejecta in either CCSNe or neutron star mergers. Consequently, our study is 
complementary to post-processing studies based on specific simulations 
of these events. With regard to the $p$-isotopes, we find that proton-rich winds 
can make dominant contributions to the solar abundance of $^{98}\mathrm{Ru}$, 
significant contributions to those of $^{96}$Ru ($\lesssim 40\%$) and $^{92}$Mo 
($\lesssim 27\%$), and relatively minor contributions to that of $^{94}$Mo 
($\lesssim 14\%$). In contrast, neutron-rich winds make negligible contributions 
 to the solar abundances of $^{92,94}$Mo and cannot produce $^{96,98}$Ru.
However, we have shown that some neutron-rich winds can account for 
the peculiar patterns of $^{95,97,98,100}$Mo in SiC X grains.

In conclusion, our results strongly suggest that the solar abundances 
of $^{92,94}$Mo are dominantly produced by sources other than the
neutrino-driven wind. In addition, while the wind can be a dominant source for
the $^{98}$Ru in the solar system, other sources are also required to 
account for the $^{96}$Ru.

\acknowledgments

This work was supported in part by the Helmholtz-University Young Investigator
grant No. \mbox{VH-NG-825}, Deutsche Forschungsgemeinschaft through \mbox{SFB 1245}, BMBF under grant No. 05P15RDFN1,
\mbox{ERC 677912} \mbox{EUROPIUM}, and the US DOE grant \mbox{DE-FG02-87ER40328}.  
We acknowledge useful discussions with Camilla
J. Hansen, Maria Lugaro, Marco Pignatari, and Claudia Travaglio.


\bibliographystyle{apj.bst}
\bibliography{paper}

\end{document}